# A Quantitative Analytical Model for Predicting and Optimizing the Rate Performance of Battery Cells

Fan Wang[1] and Ming Tang[1*]

1. Department of Materials Science & NanoEngineering, Rice University, Houston, TX 77005, USA.

[*] Corresponding author email: mingtang@rice.edu

**Abstract**

An important objective of designing lithium-ion rechargeable battery cells is to maximize their rate performance without compromising the energy density, which is mainly achieved through computationally expensive numerical simulations at present. Here we present a simple analytical model for predicting the rate performance of battery cells limited by electrolyte transport without any fitting parameters. It exhibits very good agreement with simulations over a wide range of discharge rate and electrode thickness and offers a speedup of $>10^5$ times. The optimal electrode properties predicted by the model are of less than 10% difference from simulation results, suggesting it as a attractive computational tool for the cell-level battery architecture design. The model also offers important insights on ways to improve the rate performance of thick electrodes, including avoiding electrode materials (e.g. $LiFePO_4$, $Li_4Ti_5O_{12}$) whose open-circuit potentials are insensitive to the state of charge and utilizing lithium metal anode to synergistically accelerate electrolyte transport within thick cathodes.

## I. Introduction

Nearly thirty years after the debut of commercial rechargeable lithium-ion batteries (LIBs), LIBs was recognized by the 2019 Nobel Prize in Chemistry as a key energy storage technology for wide-ranging applications from portable devices, electrical vehicles to grid load balancing. In the drive towards developing better batteries with lower costs, there exists a perpetual trade-off between the energy and power densities. As illustrated by the well-known Ragone plot[1], improvement in one metric usually leads to the degradation of the other. Recently, the use of thick electrodes has received significant interest as a way to enhance the gravimetric / volumetric energy of battery cells and cut materials and manufacturing costs by reducing the fraction of inactive components such as separator and current collectors[2-8]. However, a main challenge faced by thick electrodes is their inferior rate performance. Effective alleviation of this issue requires careful optimization of cell parameters for targeted applications and the development of novel electrode architectures (e.g. low tortuosity electrodes)[9-18] that can slow down the rate capability decay with increasing electrode thickness.

The current standard approach to predicting the rate performance of battery cells and optimizing their structures is numerical simulations[19-22] based on the porous electrode theory pioneered by Newman and co-workers[23,24]. Such simulations are often called pseudo-two-dimensional (P2D) simulations because they involve solving the governing equations for ionic (or electronic) transport in electrolyte (or solid phase) along a macroscopic length scale, which are coupled with the lithium diffusion equation within electrode particles along a microscopic length scale. While the P2D model provides a comprehensive description of the kinetic processes during (dis)charge, it is also computationally intensive to solve. Various reformulation and reduced order modeling techniques[25-28] have been developed to accelerate P2D simulations, but the computational burden is considerable for battery cell optimization, which requires a large number of objective function evaluations and a time-dependent simulation for each evaluation.

In contrast to P2D simulations, (semi-)empirical battery models such as equivalent-circuit models[29,30] are simple and fast to execute and widely used in battery management systems. Parameters in these models need to be fitted against experimental data, e.g. from cycling or impedance measurements, and valuable insights can be obtained from the fitted parameter values[31-33]. However, it is often not easy to determine the quantitative relations between the fitting parameters and the physical properties of the systems, which makes the application of these models to the cell design and optimization not straightforward. Serving as an intermediate between the P2D and (semi-)empirical models, physics-based analytical models[34-39] have long been developed for predicting battery electrode performance. Even with the constantly improving capability of computers, this type of models are desirable because they not only are efficient to solve but also shed light on the structure-performance relation of battery cells in a more transparent way.

Analytical models are often derived by simplifying the P2D model for situation where one kinetic process is much more sluggish than the others and dominates the (dis)charge behavior. Examples include the single-particle model[38,39] for (dis)charging limited by solid-state diffusion, and the reaction zone model[35,40] that assumes ohmically dominated (dis)charge processes whereas the concentration gradients in electrolyte and electrode particles can be neglected. Nevertheless, advances in the LIB technology have rendered the limiting factors considered by these models less significant as slow solid-state diffusion can be addressed by particle size reduction and low electronic conductivity can be improved by conductive additives or coatings. Compelling evidence shows that the rate performance of LIBs based on today's commercial electrode materials and liquid electrolytes is mainly limited by electrolyte transport[3,5,41-44], which builds up a large salt concentration gradient in electrolyte during cycling. This limitation has become a major impediment to the on-going push for fast charging and the development of thick battery electrodes[10,11]. Despite its technological relevance, there are few efforts in deriving simplified analytical models for (dis)charging in the electrolyte-limited regime, partly because great complexity arises from the nonlinear coupling between the electrolyte transport, reaction flux, electrical potentials and state of charge (SOC) of electrode material in the governing equations of the porous electrode theory. While a number of previous studies[5,39,42] provide valuable insights in the discharge performance in this regime, predictions by these models are qualitative and do not replace the P2D simulations.

In this work, we present a quantitative analytical model for predicting the (dis)charge behavior of battery cells limited by electrolyte transport. We developed the model by decoupling the governing equations of the porous electrode theory by making reasonable assumptions of the electrolyte transport and electrode reaction behavior generalized from P2D simulations. The model gives analytical expressions of the galvanostatic discharge capacity of both half and full cells. Without any fitting parameters, the model exhibits very good agreement with P2D simulations over a wide range of electrode thickness and discharge rate, which can be applied to battery cell optimization with >$10^5$-fold speedup compared to P2D simulations. Furthermore, simple scaling relations derived from the model offer valuable insights on the dependence of discharge performance on the electrode properties, including: i) electrode materials that exhibit a strong SOC dependence of open-circuit potentials such as NMC and graphite have intrinsically better rate performance than those that do not (e.g. LiFePO$_4$ and Li$_4$Ti$_5$O$_{12}$) in the electrolyte-transport-limited regime and are preferred for thick electrode applications, and ii) thick cathodes have significantly better rate capability when paired with Li metal than conventional intercalation anodes (e.g. graphite or Li$_4$Ti$_5$O$_{12}$), which points to a clear synergy between the development of the Li metal anode and the wider adoption of thick-electrode batteries.

## II. Results and Discussion

### II.1 Electrolyte transport and electrode reaction behavior

The analytical model is developed based on two key observations from P2D simulations, i.e. i) electrolyte transport can attain a (pseudo-)steady state during discharge, which allows the time-dependent simulations to be reduced to a stationary problem, and ii) the electrode reaction flux distribution can be categorized into two types of distinct behavior exhibited by common electrode materials. The first type of electrode materials include the mainstream layered oxide cathodes NMC and NCA, which are characterized by a strong dependence of their equilibrium (or open-circuit) potentials $U_{eq}$ on SOC. The second type of electrode compounds, exemplified by LiFePO$_4$ (LFP) and Li$_4$Ti$_5$O$_{12}$ (LTO), feature SOC-independent $U_{eq}$ due to prominent first-order phase transition(s) upon (dis)charging. Below we use representative P2D simulations to elucidate these two types of discharging behavior and motivate the simplifying assumptions employed in the analytical model.

*Type 1: "Uniform-reaction" electrodes.* Figure 1a-c presents the discharge process of an NMC111 half cell, which refers to battery cells consisting of a cathode and Li metal anode. The cathode is 250-μm thick and discharged at a current density 20mA/cm$^2$ or 1.5C, and electrolyte is 1M LiPF$_6$ in EC:DMC (50:50 wt%). Other parameters employed in the simulation are listed in Table S1 in Supplementary Information (SI). Figure 1a shows that salt depletion ($c = 0$) occurs in electrolyte near the current collector shortly after discharging starts. After an initial transient period of <10% depth of discharge (DoD), salt concentration establishes a steady state distribution in electrolyte, which persists up to ~50% DoD before discharging terminates at 62% DoD. During the steady-state period, the cathode region can be divided into a salt *depletion zone* (DZ), in which salt concentration $c$ is nearly zero, and the complementary *penetration zone* (PZ). Figure 1b shows that the Li intercalation flux is close zero in DZ but a nearly constant flux is present inside PZ. When discharging is terminated at the cut-off voltage (3 V) at DoD ≈ 62%, the majority of electrode particles in PZ (DZ) are close to be fully discharged (charged) except for a transition region near the DZ/PZ boundary, see Figure 1c. Figure S1a-c in SI shows that NMC cathode in a full cell with graphite (Gr) anode displays similar characteristics. Such discharging behavior may be approximated by a simplified "*uniform reaction*" (UR) model (Figure 1d):

i) Electrolyte transport maintains a steady state throughout the discharging process.

ii) Electrode reaction flux is zero in the salt DZ and has a uniform value in the salt PZ.

iii) Discharging ends when electrode particles in PZ are fully discharged.

We emphasize that here "uniform reaction" implies a homogeneous reaction flux within PZ instead of the whole electrode. Doyle and Newman (DN) previously developed an analytical model that also assumes a uniform reaction distribution[39] with two major differences from ours. The DN model assumes a uniform reaction flux spanning the entire cathode and that the discharge process is terminated as soon as salt depletion occurs at the cathode/current-collector interface. It thus only accounts for the discharge capacity obtained during the transient stage before the establishment of steady-state electrolyte transport. As simulations show, discharge continues after the onset of salt depletion and the steady-state stage could be responsible for the majority of the discharge capacity.

*Type 2: "Moving-zone-reaction" electrodes.* Figure 1e-g demonstrates the discharge behavior of an LFP half cell with a 250μm-thick cathode and discharged at 20mA/cm$^2$ or 1.8C. Other simulation parameters are listed in Table S1 in SI. Unlike NMC111, Li intercalation in the LFP electrode is characterized by a moving reaction front. Figure 1f shows that a sharp reaction flux peak forms near the separator and travels towards the current collector upon discharging. As shown in Figure 1e and 1g, the reaction front divides the cathode into a PZ, in which the electrode is almost fully discharged and salt concentration maintains a linear profile, and an incomplete salt DZ, in which salt is partially depleted and electrode particles are barely reacted. PZ expands at the expense of DZ, which causes the salt concentration in DZ to drop continuously. When DZ becomes completely depleted in salt, PZ can no longer expand and discharging is terminated. Electrolyte transport reaches a pseudo-steady state at the end of discharge, when the time derivative of salt concentration momentarily vanishes. Figure S1d-f in SI shows that LFP in a full cell with Gr anode displays similar characteristics. The observed discharging behavior can be idealized into a simple "*moving-zone reaction*" (MZR) model (Figure 1h):

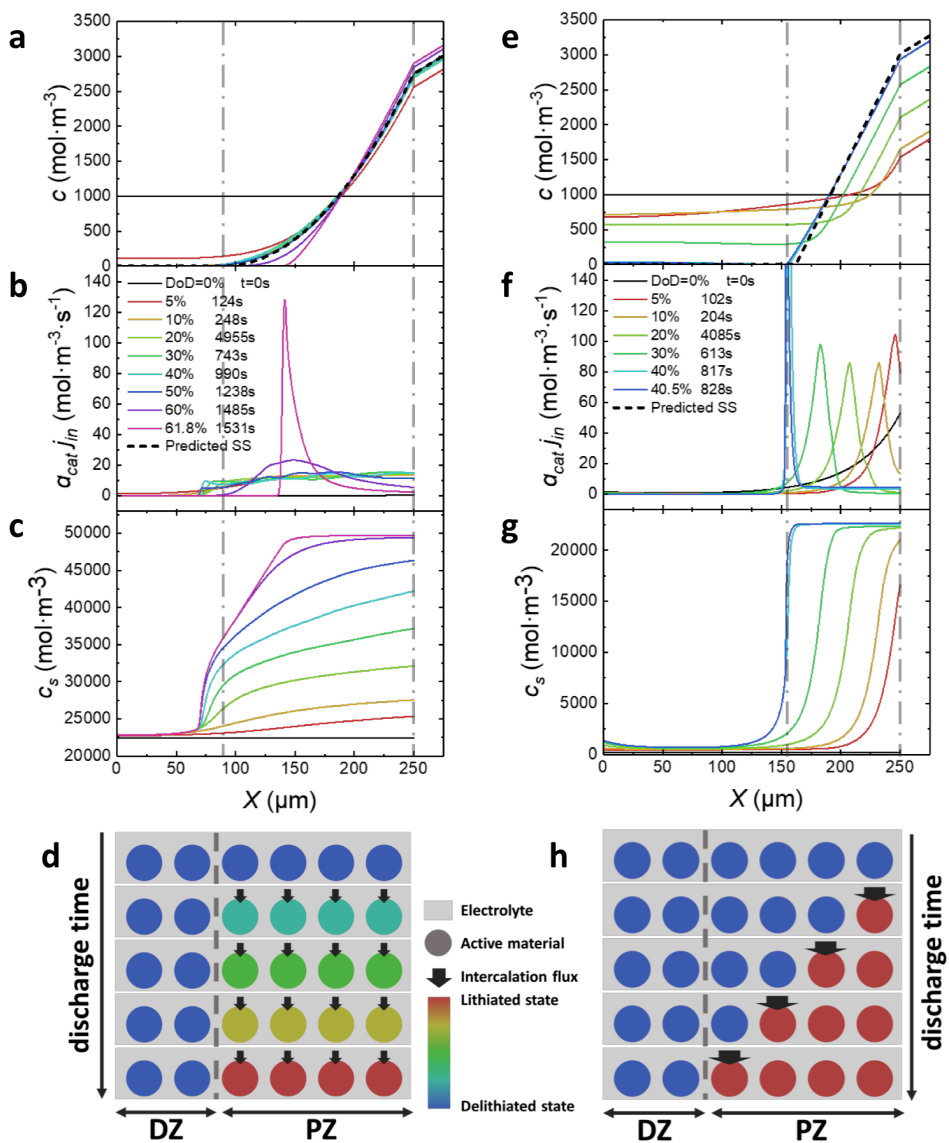

**Figure 1. Discharge behavior of NMC and LFP half cells predicted by P2D simulations.** The salt concentration $c$, reaction flux $a_i j_{in}$ and the average Li concentration in cathode particles $c_S$ at different DoDs are shown in **a-c** for NMC half cell and **e-g** for LFP half cell, respectively. $X = 0$ is at the cathode/current-collector interface. The dashed lines in **a** and **e** represent the (pseudo-)steady-state electrolyte distribution predicted by the analytical model. **d and h**, idealized uniform-reaction (UR) and moving-zone-reaction (MZR) behavior of electrode materials, respectively.

i) Li intercalation occurs at an infinitely sharp, moving reaction front separating salt PZ and DZ.

ii) Electrode particles in the PZ (DZ) are fully discharged (charged).

iii) Discharging ends when salt is completely depleted in DZ and electrolyte transport reaches a pseudo-steady state.

The MZR behavior is analogous to the reaction zone model proposed by Tiedemann and Newman for ohmically limited electrodes[35,40], which is also considered by Doyle and Newman[39]. However, there is a lack of discussion in literature on what electrodes should exhibit UR or MZR characteristics and the origin of such distinction, which is rooted in the intrinsic thermodynamic properties of electrode materials, specifically the SOC dependence of $U_{eq}$. Reaction flux at the electrode surface is often described by the Butler-Volmer equation,

$$j_{in} = i_0 \left[ \exp(\alpha F \eta / RT) - \exp(-(1-\alpha) F \eta / RT) \right] / F, \qquad 1)$$

$j_{in}$ is controlled by the surface overpotential $\eta \equiv \Phi_L - \Phi_S + U_{eq}(\text{SOC}_s)$, where $\text{SOC}_s$ is the local SOC at the particle surface. Suppose that SOC is uniform at the beginning of discharge. The ionic / electronic resistances within a cell result in spatially varied $\Phi_L - \Phi_S$ and initially non-uniform $j_{in}$ as illustrated in Figure 2a. If $U_{eq}$ has a strong SOC dependence like NMC, however, the electrode can rectify the reaction inhomogeneity as electrode particles in regions with higher $j_{in}$ experience a larger decrease in local SOC, leading to a larger drop in $U_{eq}$ and hence $j_{in}$. This self-regulating mechanism causes the spatial gradient in $j_{in}$ to continuously decrease until a uniform reaction distribution is reached within PZ (Figure 2b), where Li intercalation is permitted. On the other hand, compounds like LFP have SOC-independent $U_{eq}$, which cannot compensate the gradient of $\Phi_L - \Phi_S$ to homogenize the reaction flux (Figure 2c). In the electrolyte-limited regime, reaction in these electrodes will preferentially occur at the separator, where $\eta$ is the largest, and then propagate across the electrodes after electrode particles near the separator are fully intercalated.

The reaction behavior of anode in a full cell can also be described as the UR or MZR type. The discharge simulations of NMC/LTO and LFP/LTO full cells are presented in Figure S2 in SI. Like LFP, LTO has a flat $U_{eq}(\text{SOC})$ curve. Li deintercalation within LTO also proceeds through a moving reaction front. Compared to LTO, the discharging behavior of Gr anode is somewhat more complex and displays both UR and MZR features. Graphite's $U_{eq}$ has a pronounced plateau at ~0.05 V due to the $LiC_6 \rightarrow LiC_{12}$ staging transition. Accordingly, Gr exhibits the MZR behavior at the early discharge stage, where a reaction flux peak can be seen traveling across the anode (Figure S1b,e in SI). After DoD reaches >30%, however, the anode's $U_{eq}$ rises above the plateau and its reaction behavior switches to the UR type. Since there is no salt depletion within anode upon discharging, the reaction flux is uniformly distributed over the entire anode region.

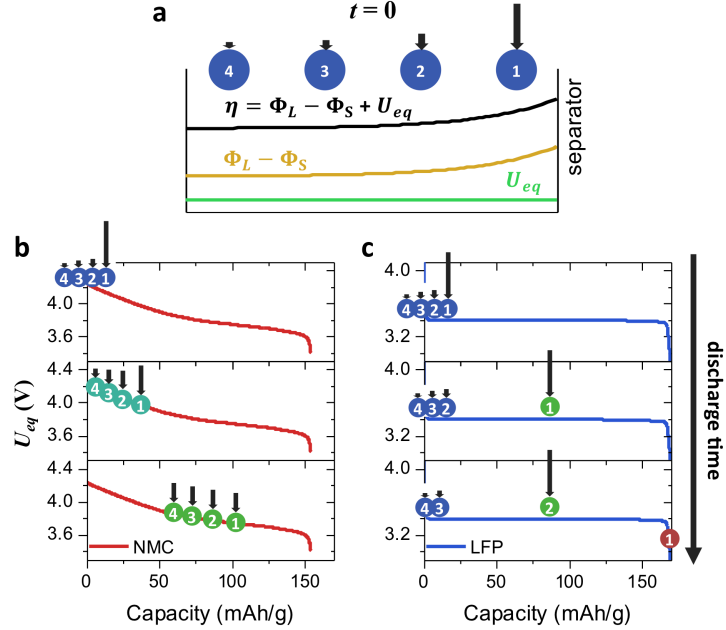

**Figure 2. Thermodynamic origin of the UR vs MZR behavior of electrode materials. a.** Schematic of the spatial distributions of the equilibrium potential $U_{eq}$, electrolyte / electrode potential difference $\Phi_L - \Phi_S$ and surface overpotential $\eta$ inside the cathode at the beginning of the discharge process. **b** and **c** Schematics of the evolution of the reaction flux and SOC distributions in UR-type cathodes (**b**) and MZR-type cathodes (**c**) upon discharging. Arrows above the particles indicate the magnitude of the reaction flux.

## II.2 Analytical model

The simplified UR and MZR behavior described in the last section allows for analytical predictions of discharge performance. Because the width of salt PZ, or the penetration depth $L_{PZ}$, is central to the rate performance of the electrolyte-limited discharge process, we set to obtain an analytical expression of $L_{PZ}$ as a function of discharge current, electrode, electrolyte and separator properties.

We start from the mass balance and current continuity equations of a binary electrolyte in the porous electrode theory[22,23]:

$$\epsilon_i \frac{\partial c}{\partial t} = \nabla \cdot \left( \frac{\epsilon_i}{\tau_i} D_{amb} \nabla c \right) + \nabla \cdot \left( \frac{(1-t_+)\vec{i}}{F} \right) \qquad 2)$$

$$\nabla \cdot \vec{i} = -F a_i j_{in} \qquad 3)$$

where subscript $i$ represents cathode ($i = cat$) or separator ($i = sep$). With the assumption of (pseudo-)steady-state electrolyte transport, Eq. 2 becomes:

$$\frac{\partial}{\partial x}\left(\frac{\epsilon_i}{\tau_i}D_{amb}\frac{\partial c}{\partial x}\right) - a_i j_{in}(1-t_+) = 0 \qquad (4)$$

in which a constant $t_+$ and homogeneous in-plane $c$ are assumed. Inside PZ, $j_{in}$ is equal to $I/Fa_{cat}L_{PZ}$ for UR and 0 for MZR electrodes, respectively. Since $c$ is 0 in DZ, Eq. 4 only needs to be solved within PZ and separator with the following boundary conditions at the PZ/DZ interface,

$$c = 0 \qquad (5)$$

$$\frac{\epsilon_{cat}D_{amb}}{\tau_{cat}}\frac{\partial c}{\partial x} = \begin{cases} 0 & \text{UR} \\ I(1-t_+)/F & \text{MZR} \end{cases} \qquad (6)$$

Setting $x = 0$ at the cathode/current-collector interface and treating $D_{amb}$ as constant, the solution to Eqs. 4-6 is given by

$$\text{UR:} \quad c = \begin{cases} 0 & 0 \leq x < L_{cat} - L_{PZ} & \text{(DZ)} \\ \dfrac{t_{cat}}{\epsilon_{cat}}\dfrac{I(1-t_+)}{2FD_{amb}L_{PZ}}(x - L_{cat} + L_{PZ})^2 & L_{cat} - L_{PZ} \leq x < L_{cat} & \text{(PZ)} \\ \dfrac{t_{sep}}{\epsilon_{sep}}\dfrac{I(1-t_+)}{FD_{amb}}(x - L_{cat}) + \dfrac{t_{cat}}{\epsilon_{cat}}\dfrac{I(1-t_+)}{2FD_{amb}}L_{PZ} & L_{cat} \leq x < L_{cat} + L_{sep} & \text{(separator)} \end{cases} \qquad (7)$$

$$\text{MZR:} \quad c = \begin{cases} 0 & 0 \leq x < L_{cat} - L_{PZ} & \text{(DZ)} \\ \dfrac{t_{cat}}{\epsilon_{cat}}\dfrac{I(1-t_+)}{FD_{amb}}(x - L_{cat} + L_{PZ}) & L_{cat} - L_{PZ} \leq x < L_{cat} & \text{(PZ)} \\ \dfrac{t_{sep}}{\epsilon_{sep}}\dfrac{I(1-t_+)}{FD_{amb}}(x - L_{cat}) + \dfrac{t_{cat}}{\epsilon_{cat}}\dfrac{I(1-t_+)}{FD_{amb}}L_{PZ} & L_{cat} \leq x < L_{cat} + L_{sep} & \text{(separator)} \end{cases} \qquad (8)$$

For half cells, the unknown $L_{PZ}$ in the solution is determined by substituting Eq. 7 and 8 into the salt conservation equation

$$\int_0^{L_{cat}} \epsilon_{cat} c\, dx + \int_{L_{cat}}^{L_{cat}+L_{sep}} \epsilon_{sep} c\, dx = (\epsilon_{cat}L_{cat} + \epsilon_{sep}L_{sep})c_0 \qquad (9)$$

The obtained $L_{PZ}$ expressions for half cells with UR and MZR cathodes are listed in Eqs. 10 and 11, respectively.

To derive $L_{PZ}$ for full cells, Eq. 4 needs to be extended to the anode region. Although Gr anode exhibits the hybrid reaction behavior, it develops a homogeneous reaction distribution at the later discharging stage (Figure S1 in SI) and so a uniform deintercalation flux $j_{in} = -I/(Fa_{an}L_{an})$ can be assumed for the entire anode region when DoD$_f$ is not too small. See Supplementary Note S1 for the detailed derivation. The obtained $L_{PZ}$ expressions for UR- and MZR-cathode/Gr full cells are given in Eqs. 12 and 13, respectively. The expressions of $L_{PZ}$ for full cells containing MZR-type anodes like LTO are also derived in Supplementary Note S1.

UR-cathode/Li half cell:
$$L_{PZ} = -\frac{3\epsilon_{sep}}{2\epsilon_{cat}}L_{sep} + \sqrt{\frac{6FD_{amb}c_0}{\tau_{cat}I(1-t_+)}(\epsilon_{cat}L_{cat}+\epsilon_{sep}L_{sep})+\left(\frac{9\epsilon_{sep}^2}{4\epsilon_{cat}^2}-\frac{3\tau_{sep}}{\tau_{cat}}\right)L_{sep}^2} \qquad 10)$$

MZR-cathode/Li half cell:
$$L_{PZ} = -\frac{\epsilon_{sep}}{\epsilon_{cat}}L_{sep} + \sqrt{\frac{2FD_{amb}c_0}{\tau_{cat}I(1-t_+)}(\epsilon_{cat}L_{cat}+\epsilon_{sep}L_{sep})+\left(\frac{\epsilon_{sep}^2}{\epsilon_{cat}^2}-\frac{\tau_{sep}}{\tau_{cat}}\right)L_{sep}^2} \qquad 11)$$

UR-cathode/Gr full cell:
$$L_{PZ} = -\frac{3(\epsilon_{sep}L_{sep}+\epsilon_{an}L_{an})}{2\epsilon_{cat}}$$
$$+\sqrt{\frac{6FD_{amb}c_0}{\tau_{cat}I(1-t_+)}(\epsilon_{cat}L_{cat}+\epsilon_{sep}L_{sep}+\epsilon_{an}L_{an})+\left(\frac{9\epsilon_{sep}^2}{4\epsilon_{cat}^2}-\frac{3\tau_{sep}}{\tau_{cat}}\right)L_{sep}^2+\left(\frac{9\epsilon_{sep}\epsilon_{an}}{2\epsilon_{cat}^2}-\frac{6\epsilon_{an}\tau_{sep}}{\epsilon_{sep}\tau_{cat}}\right)L_{sep}L_{an}+\left(\frac{9\epsilon_{an}^2}{4\epsilon_{cat}^2}-\frac{2\tau_{an}}{\tau_{cat}}\right)L_{an}^2} \qquad 12)$$

MZR-cathode/Gr full cell:
$$L_{PZ} = -\frac{\epsilon_{sep}L_{sep}+\epsilon_{an}L_{an}}{\epsilon_{cat}}$$
$$+\sqrt{\frac{2FD_{amb}c_0}{\tau_{cat}I(1-t_+)}(\epsilon_{cat}L_{cat}+\epsilon_{sep}L_{sep}+\epsilon_{an}L_{an})+\left(\frac{\epsilon_{sep}^2}{\epsilon_{cat}^2}-\frac{\tau_{sep}}{\tau_{cat}}\right)L_{sep}^2+2\left(\frac{\epsilon_{sep}\epsilon_{an}}{\epsilon_{cat}^2}-\frac{\epsilon_{an}\tau_{sep}}{\epsilon_{sep}\tau_{cat}}\right)L_{sep}L_{an}+\left(\frac{\epsilon_{an}^2}{\epsilon_{cat}^2}-\frac{2\tau_{an}}{3\tau_{cat}}\right)L_{an}^2} \qquad 13)$$

The normalized discharge capacity $DoD_f$ can be estimated by the ratio of $L_{PZ}$ to the cathode thickness $L_{cat}$. When $L_{PZ} > L_{cat}$, $DoD_f$ should be set to 1 since this means that electrolyte transport does not limits the discharge capacity at all. The model may give negative $L_{PZ}$ for full cells with large electrode thickness and discharge rates, in which case $DoD_f = 0$ is assigned. Therefore,

$$DoD_f = \begin{cases} 1 & L_{PZ} > L_{cat} \\ L_{PZ}/L_{cat} & 0 \le L_{PZ} \le L_{cat} \\ 0 & L_{PZ} < 0 \end{cases} \qquad 14)$$

In addition to predicting the discharge capacity, the model can also reveal other useful quantities of the discharge process such as the potential drops across the electrodes and separator and the overall energy efficiency, which can be calculated from the electrolyte concentration profiles given by Eqs. 7 and 8 together with Eqs. S1 and S2 in SI. With minor modification, the model is also applicable to the galvanostatic charging process limited by electrolyte transport. In practical applications, however, charging usually needs to be terminated before salt depletion occurs to prevent lithium plating on anode particle surface, which is a major concern for capacity fading. In this case, our model will give less conservative predictions.

## II.3 Comparison with P2D simulations

In this section, we examine how well the derived analytical model approximates P2D simulations. Figure 1a,e and Figure S1a,d in SI show that the predicted (pseudo-)steady-state electrolyte concentration profiles (dashed lines) have excellent agreement with P2D simulations for both half and full cells. In Figure 3, the rate-dependence of $DoD_f$ calculated from the model (dashed lines) is compared with the P2D simulation results (solid symbols) for NMC and LFP half / full cells, which again agree very well with each other over a wide range of C rates and electrode thickness values. Notably, Figure 3 reveals the existence of a critical C-rate $C_{crit}$ for each type of cell configuration, which represents the C rate above which $DoD_f$ starts to drop below 100% due to sluggish electrolyte transport. Nevertheless, the discrepancy between model predictions and simulations becomes more significant for full cells at large $L_{cat}$ (250 and 300 μm) and high rates where $DoD_f < 0.3$. The reasons are two-fold. First, Gr anode displays MZR behavior at small DoD, which deviates from the uniform reaction assumption employed in Eqs. 12 and 13. Second, discharging in these cases is prematurely terminated before electrolyte transport establishes the (pseudo-)steady state. Figure S3 in SI compares the analytical model with simulations for NMC/LTO and LFP/LTO full cells, which also show very good agreement.

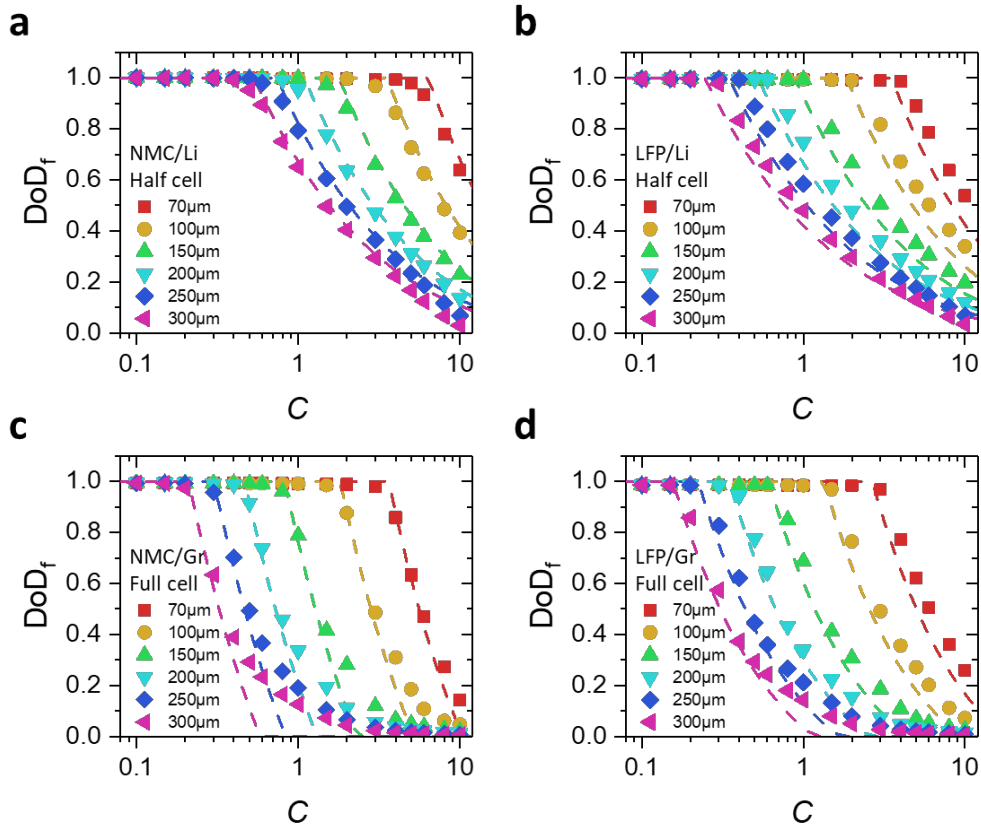

**Figure 3. Rate performance of battery cells predicted by P2D simulations (symbols) vs. the analytical model (dashed lines). a** NMC half cells. **b** LFP half cells. **c** NMC/Gr full cells. **d** LFP/Gr full cells. Different symbols and line colors correspond to different $L_{cat}$ values between 70 μm and 300 μm. Anode thickness $L_{an}$ is matched to $L_{cat}$ to give an anode/capacity capacity ratio of 1.18 in full cells.

To comprehensively test the model fidelity, 246 P2D simulations were performed for each type of cathode (NMC vs LFP) and cell configuration (half vs full) over a large range of discharge rates (0.1C – 10C), electrode thickness (70 – 300 μm) and other cell properties. The simulated DoD$_f$ is compared against $L_{PZ}/L_{cat}$ predicted by the model in Figure 4a-d. Overall, the model provides a satisfactory approximation to P2D simulations. In the case of half cells, >94% of the model predictions have a relative error of less than 10% (or 20%), and the average error is 5.1% (or 8.1%) for NMC (or LFP), see Figure S4a,b in SI. For NMC/Gr and LFP/Gr full cells, model predictions show notable difference from simulations at DoD$_f$ < 0.3 for reasons discussed above. If excluding the low DoD$_f$ cases, however, the model compares very favorably with simulations and demonstrates similar accuracy as seen in half cells (Figure S4c,d in SI). Perhaps more importantly, the model reliably predicts $C_{crit}$ for all of the cell configurations with an average relative error of 9.6% as shown in Figure 4e,f. Accurate assessment of $C_{crit}$ is critical for battery cell design as it informs the acceptable cycling conditions to avoid inferior performance.

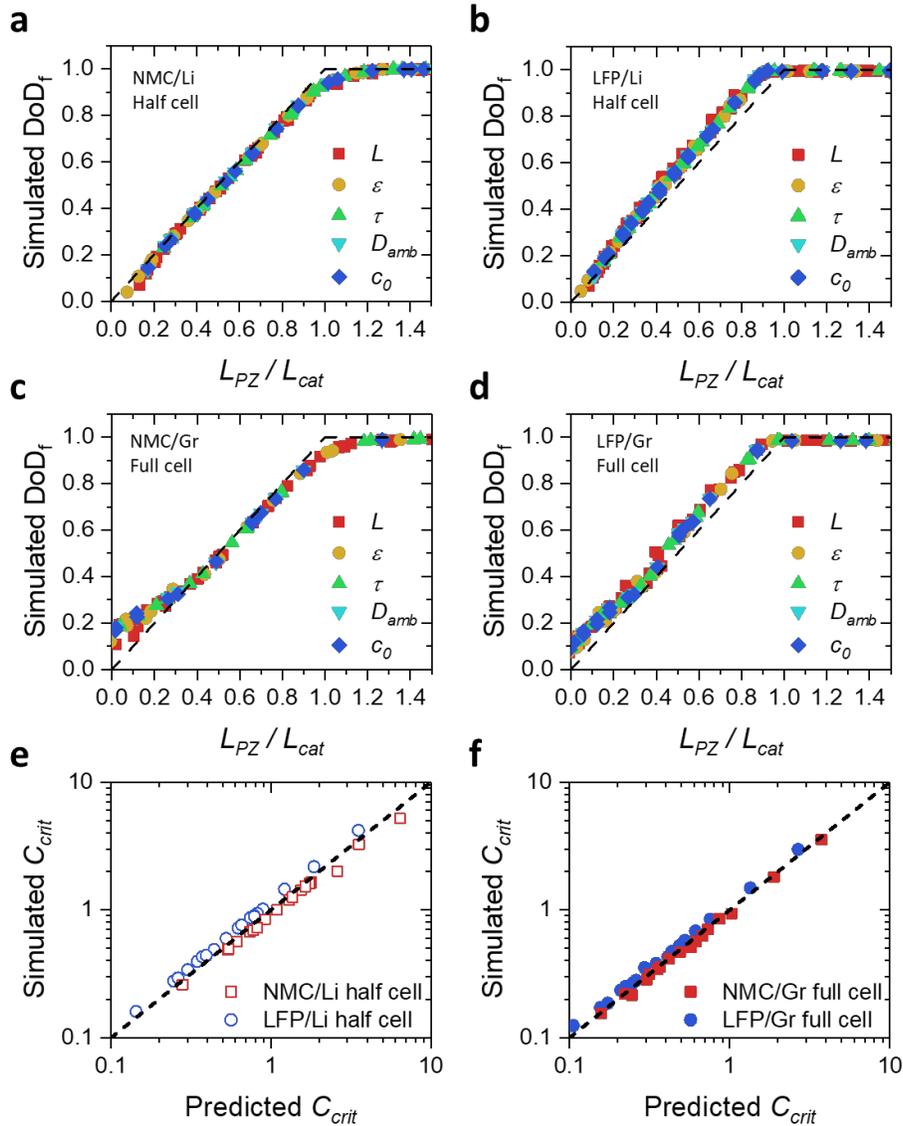

**Figure 4. Comparison between the analytical model and P2D simulations. a-d** Normalized discharge capacity DoD$_f$ from simulations vs $L_{PZ}/L_{cat}$ predicted by the model for NMC half cells (**a**), LFP half cells (**b**), NMC/Gr full cells (**c**) and LFP/Gr full cells (**d**). Dashed lines are the DoD$_f$ ~ $L_{PZ}/L_{cat}$ relation

predicted by Eq. 14. Each plot includes five sets of simulations marked by different symbols. In one set of simulations, $L_{cat}$ takes several values between 70 $\mu$m and 300 $\mu$m while the C rate varies between 0.1C and 10C. In each of the other four sets of simulations, one of the four parameters ($\epsilon_{cat}$, $\tau_{cat}$, $D_{amb}$ and $c_0$) is set to 0.5×, 0.75×, 1×, 1.25× and 1.5× of its baseline value ($\epsilon_{cat}$ = 0.25, $\tau_{cat}$ = 2, $D_{amb}$ = 2.95·10$^{-10}$ m$^2$/s, and $c_0$ = 1 M) while the C rate varies between 0.1C and 5C. For full cells, $L_{an}$, $\epsilon_{an}$ or $\tau_{an}$ vary in proportion to $L_{cat}$, $\epsilon_{cat}$ or $\tau_{cat}$, respectively. **e and f** Critical C rate $C_{crit}$ determined from simulations vs model predictions for NMC and LFP half cells (**e**) and NMC/Gr and LFP/Gr full cells (**f**). $C_{crit}$ from simulations is calculated by extrapolating data to DoD$_f$ = 1.

## II.4 Scaling relations

With additional approximations, the expressions of $L_{PZ}$ shown in Eqs. 10-13 can be further simplified to provide a less accurate but more revealing scaling description of the relation between $L_{PZ}$ and various cell properties and discharge rate. A simpler expression of $L_{PZ}$ for half cells can be obtained by neglecting the separator thickness ($L_{sep}$ = 0) in Eqs. 10 and 11:

$$L_{PZ} = \sqrt{\gamma_h \frac{\epsilon_{cat} F D_{amb} c_0 L_{cat}}{\tau_{cat} I (1-t_+)}} \qquad \text{(half cells)} \qquad 15)$$

where $\gamma_h$ = 6 or 2 for UR or MZR cathodes, respectively.

To simplify the expression of $L_{PZ}$ for full cells with Gr anode, we start from the solution of $c(x)$ (Eqs. 7 and 8), from which one can write down:

$$L_{PZ} = \frac{\lambda \epsilon_{cat} F D_{amb}}{\tau_{cat} I (1-t_+)} c_{cat/sep} \qquad 16)$$

where $c_{cat/sep}$ is the salt concentration at the cathode/separator interface, and $\lambda$ = 2 or 1 for UR- or MZR-type cathodes, respectively. Eq. 16 is in fact valid for both full and half cells. Interestingly, P2D simulations (Figure S1a,d) show that $c_{cat/sep}$ in full cells does not deviate too much from the average salt concentration $c_0$ during discharge and thus may be treated as a constant. In Supplementary Note S2, we derive an estimated value of $c_{cat/sep} \approx c_0$ (or $12c_0/11$) for UR- (or MZR-)type cathodes paired with Gr anode. This leads to an approximate $L_{PZ}$ for full cells containing Gr anode:

$$L_{PZ} \approx \gamma_f \frac{\epsilon_{cat} F D_{amb} c_0}{\tau_{cat} I (1-t_+)} \qquad \text{(full cells)} \qquad 17)$$

where $\gamma_f$ = 2 or 12/11 for UR or MZR cathodes, respectively. We note that a similar expression of the electrolyte penetration depth as Eq. 17 is given by Gallagher et al. in Ref. 5, in which the proportionality factor $\gamma_f$ is numerically determined to be ~1.8 by fitting to P2D simulations of NCA/Gr full cells. Since NCA is a UR-type electrode, our model naturally explains this result and further the generality of the $\gamma_f$ value is not limited to NCA alone. In Supplementary Note S2, we show that Eq. 17 is also applicable to full cells with LTO anode with different $\gamma_f$ values: $\gamma_f =$ 12/7 (UR) or 1 (MZR).

Valuable insights can be obtained from the simple scaling relations given in Eqs. 15 and 17. First, they show that UR-type cathodes (e.g. NMC, NCA) with their larger $\gamma_h / \gamma_f$ values have inherently better rate capability than MZR-type cathodes like LFP in the electrolyte-transport-limited regime. This is because the presence of a uniform reaction flux in UR-type cathodes causes the salt concentration to decrease more slowly with the distance to separator, which results in a larger penetration depth. At the same cathode thickness and C rate, NMC's DoD$_f$ is 1.7-2 times of LFP in both half and full cells. The reaction behavior of the anode has a similar effect on the discharge performance, where LTO has a smaller $\gamma_f$ than Gr. In addition to the poorer rate performance, MZR-type electrodes are subject to high local reaction flux during (dis)charging, which makes them prone to excessive stress concentration and localized heat generation that will accelerate battery degradation. Therefore, UR-type electrode materials are more advantageous than MZR-type electrodes in thick electrode applications.

Second, Eqs. 15 and 17 reveal that half cells have intrinsically better rate capability than full cells, with $L_{PZ} / L_{cat}$ scaling with $(IL_{cat})^{-1/2}$ or $(CL_{cat}^2)^{-1/2}$ in the former vs $(IL_{cat})^{-1}$ or $(CL_{cat}^2)^{-1}$ in the latter. The different scaling relations suggest that DoD$_f$ decays more rapidly with $L_{cat}$ and $I$ (or $C$ rate) in full cells for identical cathodes. This is clearly seen in P2D simulations (Figure 3) especially at large electrode thickness. Eq. 16, which relates $L_{PZ}$ to $c_{cat/sep}$, explains such difference. While $c_{cat/sep}$ remains close to $c_0$ in full cells during discharging, it rises significantly above $c_0$ in half cells (Figure 1a,e). This is because the development of salt depletion inside the cathode pushes salts towards the anode side because of the electrolyte conservation. While the surplus salt can be hosted within the porous anode in full cells, it can only be accumulated within the separator in half cells. A higher salt concentration is thus built up at the cathode/separator interface in half cells, which allows the salt to penetrate deeper into the cathode to have a larger $L_{PZ}$.

Because $L_{sep} = 0$ is assumed in Eqs. 15 and 17, the simple scaling relations tend to overestimate $L_{PZ}$ since a separator of finite thickness will further slow down electrolyte transport. The scaling exponents predicted by Eqs. 15 and 17, $n = -1/2$ for hall cells and -1 for full cells, hence represent an upper limit of the actual exponent values. This is illustrated in Figure 5a,b, which plots DoD$_f$ from P2D simulations presented in Figure 3 against $CL_{cat}^2$ in logarithmic scale. It shows that $n$ is very close to -1/2 (or -1) in half (or full) cells near DoD$_f = 1$ but decreases as DoD$_f$ (or $L_{PZ}/L_{cat}$) is reduced. Eqs. 15 and 17 are thus most accurate when DoD$_f$ is relatively high or salt depletion is not severe. Figure 5a also shows that $n$ is always larger than -1 in half cells. This is because salt accumulation in the separator causes $c_{cat/sep}$ to increase with $IL_{cat}$ in half cells, which leads to a superlinear scaling between $L_{PZ}/L_{cat}$ and $1/IL_{cat}$. Therefore, a general scaling behavior can be expressed as:

$$L_{PZ} / L_{cat} \propto (IL_{cat})^n \propto (CL_{cat}^2)^n$$
$$-1 < n < -\frac{1}{2} \text{ (half cells)} \quad \text{or} \quad n < -1 \text{ (full cells)}$$
18)

The validity of Eq. 18 is supported by existing experimental works[3,5,8-10,41] that report the dependence of discharge capacity on the discharge rate and/or electrode thickness (see Supplementary Note S3 for the criteria used in selecting the experimental data.) Figure 5c and Table S3 in SI show a clear difference in the scaling exponents of full and half cells measured from the experiments. Consistent with Eq. 18, all of the half cell data exhibit $-1 < n < -1/2$ while $n < -1$ holds for full cells.

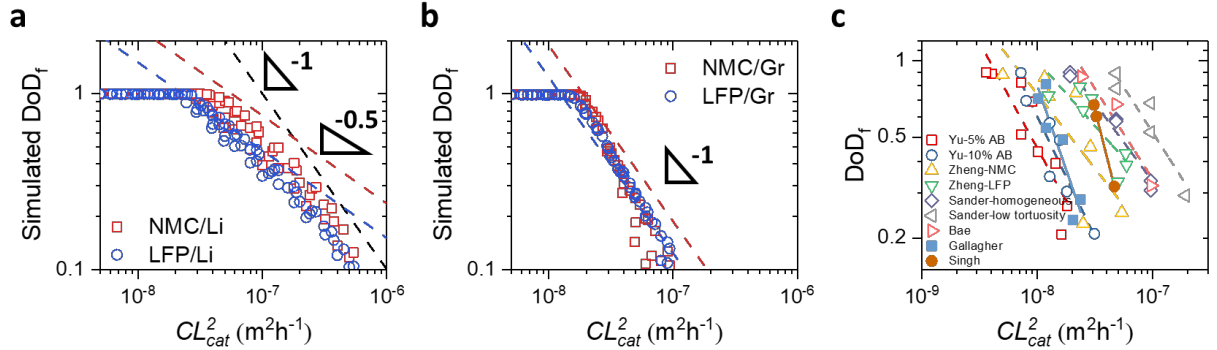

**Figure 5. Scaling relation between $DoD_f$ and $CL_{cat}^2$ displayed in P2D simulations and experiments. a and b** Simulated $DoD_f$ shown in Figure 3 is replotted against $CL_{cat}^2$ for NMC and LFP half cells (**a**) and NMC/Gr and LFP/Gr full cells (**b**). Dashed lines in **a** and **b** represent the scaling behavior predicted by Eqs. 15 and 17. **c** $DoD_f$ vs $CL_{cat}^2$ measured from experimental data in ref. [3,5,8-10,41]. Open and filled symbols represent half and full cell data, respectively, and dashed lines are linear least squares fittings. See Supplementary Note S3 in SI for criteria used in data selection.

The qualitative difference in the discharge characteristics of half vs full cells has significant practical implications. As a common practice, the rate capability of electrodes is often tested in half cells for convenience. However, our finding cautions that this approach may significantly overestimate the electrode performance in full-cell applications when discharge is kinetically limited by electrolyte transport. From another perspective, a half cell is essentially a cathode paired with a Li metal anode. Replacing conventional anodes with Li metal thus not only increases the specific/volumetric anode capacity[45] but also significantly improves the discharge performance of thick cathodes. Such a synergistic effect makes paring thick cathodes with Li metal anode a promising and perhaps necessary strategy to employ thick electrodes in high-power battery cells. This conclusion also applies to other high-capacity alloy anode materials such as silicon and its compounds. Because they require a much smaller anode thickness to match the cathode capacity, the asymmetric structure promotes salt enrichment in the separator upon discharging and improves salt penetration into the cathode.

Despite the difference, Eq. 18 suggests that $L_{cat} \propto C^{-1/2}$ in both half and full cells at a fixed $DoD_f$. In other words, cathode thickness should decrease parabolically with the $C$ rate, or conversely $C$ should decrease quadratically with $L_{cat}$, to maintain the same level of capacity utilization. This prediction explains a number of experimental observations[3,5,41], which show that such a relation is indeed obeyed in both half and full cells.

## II.5 Optimizing battery cells with the analytical model

The analytical model derived in this work provides a fast and convenient computational tool for battery cell optimization. As a demonstration, we apply it to search for the optimal electrode porosity and thickness that maximize the specific capacity of NMC/Gr full cells and NMC half cells at a given discharge rate. In the computation, $L_{cat}$ and $\epsilon_{cat}$ are parameters subject to optimization, $L_{an}$ is fixed at $1.15 L_{cat}$ and $\epsilon_{an}$ is correlated to $\epsilon_{cat}$ to give a constant anode/cathode capacity ratio of 1.18. Other parameters are the same as in the P2D simulations reported above. The properties of various cell components used in the calculation of the specific capacity are given in Table S4 in SI.

Figure 6a,b presents the calculated specific capacity $Q_w$ of NMC/Li and NMC/Gr cells at 1C discharge as a function of $L_{cat}$ and $\epsilon_{cat}$, respectively. The optimal ($L_{cat}$, $\epsilon_{cat}$) predicted by the model is marked by red circles. Each plot is generated by $10^6$ calculations, sampling 1000 $L_{cat}$ values between 50 and 600 μm and 1000 $\epsilon_{cat}$ values between 0.15 and 0.8. For accuracy benchmarking, 621 P2D simulations were carried out for each type of cell configuration in COMSOL Multiphysics® v5.3a, in which $L_{cat}$ is varied at an interval of 25 μm and $\epsilon_{cat}$ at an increment of 0.025. The simulation results are presented as square symbols in Figure 6c,d and compared to the maximal $Q_w$ as a function of $L_{cat}$ predicted by the model (dashed lines). It can be seen that the model well captures the upper limit of $Q_w$ achievable at a given cathode thickness. For better comparison, additional simulations were performed in the vicinity of the predicted optimal cathode thickness and porosity ($L_{cat}^{opt}$, $\epsilon_{cat}^{opt}$) using a smaller $L_{cat}$ interval (10 μm) to locate the global maximum of $Q_w$ more precisely. As shown in Figure 6e, the model predicts ($L_{cat}^{opt}$, $\epsilon_{cat}^{opt}$) of the NMC half cell at (216 μm, 0.256), which differ by less than 14% and 7% from the optimal electrode thickness (190 μm) and porosity (0.275) found by simulations. Figure 6f shows that the agreement is even better in the case of NMC/Gr full cells, where the model prediction ($L_{cat}^{opt}$, $\epsilon_{cat}^{opt}$) = (108.1 μm, 0.186) is less than 9% different from the simulation result at (100 μm, 0.2). Remarkably, even at a much higher 5C rate, the model can still predict ($L_{cat}^{opt}$, $\epsilon_{cat}^{opt}$) with <8% relative error compared to simulations, see Figure S6 in SI. Notably, $L_{cat}^{opt}$ of the NMC half cells is about twice of that of NMC/Gr full cells at both 1C and 5C, which again demonstrates the synergy between Li metal anode and thick cathodes for high power applications.

While the analytical model is capable of making reliable predictions, it offers enormous computational saving over P2D simulations. For simulations performed in this work, it took an average of 236 s to complete a discharge simulation in COMSOL run on a workstation (DELL Precision 5820 with Intel Xeon W-2155 processor and 128 GB memory). On the other hand, it requires only 300 ms to complete $10^6$ evaluations of the model in MATLAB on a laptop (2.2 GHz Intel Core i5 processor, 8 GB RAM), which amounts to a speedup of $10^9$ times per evaluation over P2D. Even considering the acceleration of P2D simulations via state-of-the-art techniques[25-28], the improvement in the computational efficiency achieved by the analytical model remains substantial. For example, ref. 28 reports that the running time of a P2D simulation can be reduced to 46–156 ms by using a combination of coordinate transformation, orthogonal collocation and model reformulation, but the analytical model is still $10^5$ times faster in comparison.

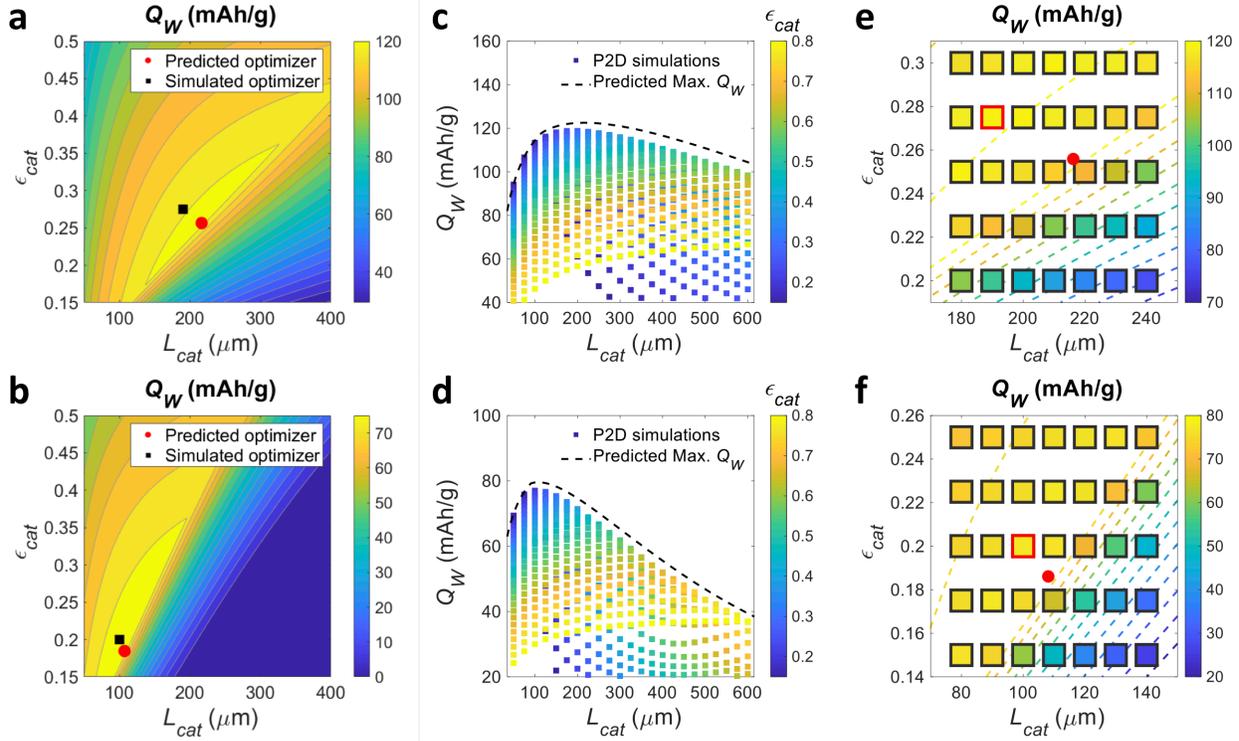

**Figure 6. Optimization of NMC half cells and NMC/Gr full cells using the analytical model. a and b** contour plots of cell-level specific capacity $Q_w$ at 1C discharge as a function of $L_{cat}$ and $\epsilon_{cat}$ for NMC half and NMC/Gr full cells, respectively. The anode/cathode capacity ratio is held constant at 1.18, and $L_{an}$ and $\epsilon_{an}$ vary in proportion to $L_{cat}$ and $\epsilon_{cat}$, respectively. The red circle and black square represent the global optimum $(L_{cat}^{opt}, \epsilon_{cat}^{opt})$ predicted by the model and P2D simulations, respectively. **c and d** Square symbols represent P2D simulations of NMC half cells (**c**) and NMC/Gr cells (**d**) at different $L_{cat}$ and $\epsilon_{cat}$. Black dashed line is the maximum $Q_w$ at each $L_{cat}$ predicted by the analytical model. **e and f** close-up of the neighborhood of $(L_{cat}^{opt}, \epsilon_{cat}^{opt})$ for 1C discharge in the design parameter space for NMC half and NMC/Gr full cells, respectively. Each square symbol represents a simulation and its filled color corresponds to simulated $Q_w$. The square symbol with red edges represents $(L_{cat}^{opt}, \epsilon_{cat}^{opt})$ located by P2D simulations, and the red circle is the model prediction.

Together with its ease of implementation, the model offers a useful alternative and supplement to P2D simulations for battery performance prediction and cell design in the electrolyte-limited regime. It could be widely utilized by battery researchers and engineers without the need for numerical simulation experiences. The method is especially powerful for exploring novel battery electrode architectures such as heterogeneous or graded electrodes, which involve a large number of design variables. One may further combine the model with P2D simulations into an efficient hybrid optimization scheme, in which the model is first used to rapidly locate the approximate global optima that are next refined by P2D simulations. This is already showcased in Figure 6e,f, where only a few tens of simulations are needed in the neighborhood of the model-predicted ($L_{cat}^{opt}$, $\epsilon_{cat}^{opt}$) to identify the optimal configurations, as against performing a much larger number of simulations over the broader parameter space. By leveraging the speed of the analytical model to scan different regions of the parameter space, this approach can also prevent the search from being trapped by local optima, which is a common pitfall of gradient-based optimization algorithms.

## III. Conclusions

In conclusion, an analytically solvable model is developed in this work to predict the rate performance of battery cells controlled by electrolyte transport, which is the dominant kinetic limiting factor in battery (dis)charging. The model simplifies the governing equations of the porous electrode theory by assuming (pseudo-)steady-state electrolyte transport and two types of electrode reaction distributions (UR vs MZR), which apply to electrode materials with (e.g. NMC) and without (e.g. LFP) a strong SOC dependence of open-circuit potentials, respectively. We derive analytical expressions of the galvanostatic discharge capacity for different combinations of electrodes and cell configurations (Eqs. 10-13, Supplementary Table S2), which exhibit very good agreement with P2D simulations over a wide range of electrode thickness and discharge rates. When applied to maximize the specific capacity of battery cells, the model predicts the optimal electrode thickness and porosity with <~10% difference from P2D simulations at negligible computational cost. The model also reveals important scaling relations (Eqs. 15-18) between the discharge capacity and electrode and electrolyte properties, which offers the following insights:

i) UR-type electrodes such as NMC and NCA can deliver significantly higher capacity utilization at large electrode thickness and discharge rates than MZR-type electrodes such as LFP and are more suited for thick electrode applications.

ii) Battery half cells (i.e. cathode/Li cells) have inherently better rate capability than their full cell counterparts in electrolyte-limited discharge process. Testing electrodes in the half cell form thus does not provide reliable indications of their performance in full cells.

iii) Pairing thick cathodes with Li metal not only increases the anode's specific capacity but also synergistically improves the discharge performance of thick cathodes, which is a promising strategy to enable thick electrodes for high-power applications.

The analytical model complements P2D simulations and provides an efficient computational tool for battery performance prediction, design and optimization.

**Method**

*P2D Simulations*. Detailed description of the P2D model can be found in numerous works in literature[19,20,22,24]. Here we summarize the equations employed in the simulations. The mass transport and current continuity equations of a binary electrolyte in a battery cell are given by Eqs. 2 and 3, and the current continuity equation in the solid phase is given by

$$\nabla \cdot \vec{i}_S = F a_i j_{in} \quad \quad 19)$$

where $a_i = 3(1-\epsilon_i)/r_i$ ($i = cat$ or $an$). The effective ionic current density $\vec{i}$ and electronic current density $\vec{i}_S$ are expressed as

$$\vec{i} = -\frac{\epsilon_i}{\tau_i}\kappa(c)\nabla\Phi_L - \frac{\epsilon_i}{\tau_i}\frac{RT\kappa(c)}{Fc}(2t_+ - 1)\left(1 + \frac{\partial \ln f_\pm}{\partial \ln c}\right)\nabla c \quad \quad 20)$$

$$\vec{i}_S = -\sigma_S \nabla \Phi_S \quad \quad 21)$$

where subscript $i = cat$ or $an$. The reaction flux on electrode particle surface $j_{in}$ obeys the Butler-Volmer equation (Eq. 1), in which the exchange current density $i_0$ is given by

$$i_0 = F k_0 c^{1-\alpha} c_S^\alpha (c_{S,max} - c_S)^\alpha \quad \quad 22)$$

The reaction flux on the Li metal anode surface is similarly described. Solid-state diffusion of Li in electrodes at the particle scale is approximated as a radial diffusion process in spherical particles:

$$\frac{\partial c_S}{\partial t} = \frac{1}{r^2}\frac{\partial}{\partial r}\left(r^2 D_S \frac{\partial c_S}{\partial r}\right) \quad \quad 23)$$

Eq. 23 is coupled to mass transport in electrolyte through the boundary condition at the particle surface

$$\left. D_S \frac{\partial c_S}{\partial r} \right|_{r=r_i} = j_{in} \quad \quad 24)$$

In full cell simulations, the anode/cathode capacity ratio is fixed at 1.18, which results in an anode/cathode thickness ratio of 1.15 (Gr/NMC), 0.95 (Gr/LFP), 1.43 (LTO/NMC) and 1.18 (LTO/LFP) when the electrode porosity values listed in Supplementary Table S1 are used. All of the simulations are implemented in COMSOL Multiphysics® 5.3a.

**List of Symbols**

| | |
|---|---|
| $a_{cat} / a_{an}$ | Volumetric surface area of electrode [m$^{-1}$] |
| $C$ | C rate |
| $C_{crit}$ | Critical C rate |
| $c$ | Salt concentration in electrolyte [mol·m$^{-3}$] |
| $c_0$ | Initial salt concentration [mol·m$^{-3}$] |
| $c_{cat/sep}$ | Salt concentration at cathode / separator interface [mol·m$^{-3}$] |
| $c_S$ | Li concentration in electrode particles [mol·m$^{-3}$] |
| $c_{S,max}$ | Maximum Li concentration in active materials [mol·m$^{-3}$] |
| $D_{amb}$ | Ambipolar diffusivity of electrolyte [m$^2$·s$^{-1}$] |
| $D_S$ | Li diffusivity in active materials [m$^2$·s$^{-1}$] |
| DoD | Depth of discharge |
| DoD$_f$ | Final depth of discharge or normalized discharge capacity |

| $F$ | Faraday constant (96485 C·mol$^{-1}$) |
|---|---|
| $I$ | Applied current density [A·m$^{-2}$] |
| $i_0$ | Exchange current density of active materials [A·m$^{-2}$] |
| $j_{in}$ | Reaction flux on active material surface [mol·m$^{-2}$·s$^{-1}$] |
| $k_0$ | Reaction rate constant [mol·m$^{-2}$·s$^{-1}$·(mol·m$^{-3}$)$^{-1.5}$] |
| $L_{cat}$ / $L_{sep}$ / $L_{an}$ | Cathode / separator /anode thickness [m] |
| $L_{cat}^{opt}$ / $L_{cat}^{crit}$ | Optimal / critical cathode thickness [m] |
| $L_{PZ}$ | Salt penetration depth [m] |
| $Q_V^0$ | Volumetric capacity of active materials [mAh·cm$^{-3}$] |
| $Q_A$ | Cell-level areal capacity [mAh·cm$^{-2}$] |
| $Q_w$ | Cell-level specific capacity [mAh·g$^{-1}$] |
| $Q_{A,max}$ / $Q_{w,max}$ | Maximal areal / specific capacity |
| $R$ | Gas constant (8.314 J·mol$^{-1}$·K$^{-1}$) |
| $r_{cat}$ / $r_{an}$ | Cathode / anode particle radius [m] |
| $T$ | Temperature [298 K] |
| $t_0$ | 1 hour (3600 s) |
| $t_+$ | Cation transference number in electrolyte |
| $U_{eq}$ | Equilibrium (open-circuit) potential of active material [V] |
| $w_{cc}$ | Weight of current collectors per cell [g·m$^{-2}$] |
| $1+\partial \ln f_\pm / \partial \ln c$ | Thermodynamic factor |
| $\alpha$ | Charge transfer coefficient |
| $\beta$ | Exponent in Bruggeman relation |
| $\epsilon_{cat}$ / $\epsilon_{sep}$ / $\epsilon_{an}$ | Cathode / separator /anode porosity |
| $\epsilon_{cat}^{opt}$ / $\epsilon_{cat}^{crit}$ | Optimal / critical cathode porosity |
| $\eta$ | Overpotential [V] |
| $\kappa$ | Electrolyte conductivity [S·m$^{-1}$] |
| $\rho_{cat}$ / $\rho_{sep}$ / $\rho_{an}$ / $\rho_{el}$ | Cathode / separator / anode / electrolyte material density [g·m$^{-3}$] |
| $\sigma_S$ | Solid phase conductivity [S·m$^{-1}$] |
| $\tau_{cat}$ / $\tau_{sep}$ / $\tau_{an}$ | Electrode tortuosity |
| $\Phi_L$ / $\Phi_S$ | Electrolyte / solid phase potential [V] |

**Data Availability**

The data that support the findings of this study are available from the corresponding author upon request.

**Acknowledgments**

FW and MT are supported by DOE under project number DE-SC0019111. Simulations were partially performed on supercomputers at the Texas Advanced Computing Center (TACC) at The University of Texas and the National Energy Research Scientific Computing Center, a DOE

Office of Science User Facility supported by the Office of Science of the U.S. Department of Energy under Contract No. DE-AC02-05CH11231.

**Contributions**

MT conceived the project. FW and MT performed the theoretical analysis and numerical calculations. MT and FW wrote the paper.

**Competing Interests**

The Authors declare no Competing Financial or Non-Financial Interests.

Supplementary Information for

# A Quantitative Analytical Model for Predicting and Optimizing the Rate Performance of Battery Cells

Fan Wang[1] and Ming Tang[1*]

1. Department of Materials Science & NanoEngineering, Rice University, Houston, TX 77005, USA.

*Corresponding author email: mingtang@rice.edu

**Supplementary Note S1** — Derivation of the analytical model for full cells

For full cells, the half-cell solution (Eqs. 7 or 8) to the steady-state electrolyte transport equation (Eq. 4) remains valid in the cathode and separator regions. The salt concentration distribution inside the anode depends on the reaction behavior of the anode and is derived for graphite and LTO anodes separately below.

I. Full cells with graphite (Gr) anode

Assuming that the reaction flux is uniform in Gr anode upon discharge, we apply $j_{in} = -I/(Fa_{an}L_{an})$ to Eq. 4 within the entire anode region. Using $c(x = L_{cat} + L_{sep})$ given by Eq. 7 or 8 as the boundary condition, salt concentration $c(x)$ inside the anode can be obtained:

*UR-type cathode* (e.g. NMC):

$$c(x) = -\frac{\tau_{an}}{\varepsilon_{an}}\frac{I(1-t_+)}{2FD_{amb}L_{an}}(x - L_{cat} - L_{sep} - L_{an})^2 + \frac{I(1-t_+)}{FD_{amb}}(\frac{\tau_{cat}}{2\varepsilon_{cat}}L_{PZ} + \frac{\tau_{sep}}{\varepsilon_{sep}}L_{sep} + \frac{\tau_{an}}{2\varepsilon_{an}}L_{an}) \quad \text{S1)}$$

$$L_{cat} + L_{sep} < x \leq L_{cat} + L_{sep} + L_{an}$$

*MZR-type cathode* (e.g. LFP):

$$c(x) = -\frac{\tau_{an}}{\varepsilon_{an}}\frac{I(1-t_+)}{2FD_{amb}L_{an}}(x - L_{cat} - L_{sep} - L_{an})^2 + \frac{I(1-t_+)}{FD_{amb}}(\frac{\tau_{cat}}{\varepsilon_{cat}}L_{PZ} + \frac{\tau_{sep}}{\varepsilon_{sep}}L_{sep} + \frac{\tau_{an}}{2\varepsilon_{an}}L_{an}) \quad \text{S2)}$$

$$L_{cat} + L_{sep} < x \leq L_{cat} + L_{sep} + L_{an}$$

Using Eqs. 7 and S1 for UR-type cathodes or Eqs. 8 and S2 for MZR-type cathodes, the unknown salt penetration depth $L_{PZ}$ can be solved from the salt conservation equation in a full cell:

$$\int_0^{L_{cat}} \epsilon_{cat} c\, dx + \int_{L_{cat}}^{L_{cat}+L_{sep}} \epsilon_{sep} c\, dx + \int_{L_{cat}+L_{sep}}^{L_{cat}+L_{sep}+L_{an}} \epsilon_{an} c\, dx = (\epsilon_{cat}L_{cat} + \epsilon_{sep}L_{sep} + \epsilon_{an}L_{an})c_0 \quad \text{S3)}$$

The obtained expressions of $L_{PZ}$ are given in Eqs. 12 and 13 in the main text.

II. Full cells with LTO anode

Because LTO exhibits MZR behavior, we assume that a penetration zone also exists near the separator inside the anode during discharge. All the anode particles within this zone are fully deintercalated, those outside the zone remain fully intercalated, and an infinitely sharp reaction flux is present at the boundary between these two regions but zero elsewhere. The width of the penetration zone in anode $L_{PZ,an}$ is correlated with the penetration depth $L_{PZ}$ in the cathode because of the conservation of Li in the solid phase:

$$L_{pz,an} = \frac{(1-\epsilon_{cat})(c_{Smax,cat}-c_{S0,cat})}{(1-\epsilon_{an})c_{S0,an}} L_{PZ} \qquad \text{S4)}$$

Using $c(x = L_{cat} + L_{sep})$ given by Eq. 7 or 8 as the boundary condition, one can solve Eq. 4 for the salt concentration $c(x)$ within the penetration zone of the anode region:

*UR-type cathode* (e.g. NMC):

$$c = \frac{\tau_{an}}{\epsilon_{an}} \frac{I(1-t_+)}{FD_{amb}}(x - L_{cat} - L_{sep}) + \frac{I(1-t_+)}{FD_{amb}}(\frac{\tau_{cat}}{2\epsilon_{cat}}L_{PZ} + \frac{\tau_{sep}}{\epsilon_{sep}}L_{sep}) \qquad L_{cat} + L_{sep} < x \leq L_{cat} + L_{sep} + L_{PZ,an} \qquad \text{S5)}$$

*MZR-type cathode* (e.g. LFP):

$$c = \frac{\tau_{an}}{\epsilon_{an}} \frac{I(1-t_+)}{FD_{amb}}(x - L_{cat} - L_{sep}) + \frac{I(1-t_+)}{FD_{amb}}(\frac{\tau_{cat}}{\epsilon_{cat}}L_{PZ} + \frac{\tau_{sep}}{\epsilon_{sep}}L_{sep}) \qquad L_{cat} + L_{sep} < x \leq L_{cat} + L_{sep} + L_{PZ,an} \qquad \text{S6)}$$

Additionally, we assume that $c(x)$ in the anode region is uniform outside the penetration zone ($L_{cat} + L_{sep} + L_{PZ,an} < x < L_{cat} + L_{sep} + L_{an}$). Substituting the above solution of $c(x)$ into Eq. S3 and solving for $L_{PZ}$, we obtain the analytical expressions of $L_{PZ}$ as given in Supplementary Table S2.

**Supplementary Note S2** — Derivation of approximate expressions of $c_{cat/sep}$ and $L_{PZ}$ for full cells

I. Full cells with graphite anode

To obtain an estimate of the salt concentration at the cathode/separator interface $c_{cat/sep}$, we replace the term $FD_{amb}/\tau_{cat}I(1-t_+)$ in Eqs. 12 and 13 in the main text with $L_{PZ}/\lambda\epsilon_{cat}c_{cat/sep}$ based on Eq. 16. Additionally, neglecting separator thickness and assuming identical properties (thickness, porosity and tortuosity) for the cathode and anode in Eqs. 12 and 13, the following approximate expressions of $c_{cat/sep}$ can be obtained:

$$c_{cat/sep} \approx \begin{cases} \dfrac{6f}{(f+1)(f+2)} c_0 & \text{UR-type cathode} \\ \dfrac{12f}{3f^2 + 6f + 2} c_0 & \text{MZR-type cathode} \end{cases} \quad \text{S7)}$$

where $f \equiv L_{PZ}/L_{cat}$ is equal to $DoD_f$ when $0 < L_{PZ} < L_{cat}$. The $c_{cat/sep}(f)$ functions above are plotted in Supplementary Figure S5, which shows that $c_{cat/sep}$ is close to $c_0$ when $f$ is not too small. For example, $c_{cat/sep}/c_0$ varies in the range of 1 – 0.7 (UR) and 1.09 – 0.98 (MZR) when $r$ decreases from 1 to 0.4. Supplementary Figure S5 also shows that the approximate $c_{cat/sep}$ given by Eq. S7 agrees well with the exact value of $c_{cat/sep}$. If replacing $c_{cat/sep}$ in Eq. 16 with its approximate value at $f = 1$, we obtain a simple expression of $L_{PZ}$ for full cells with Gr anode as given in Eq. 17 in the main text.

II. Full cells with LTO anode

An approximate expression of $c_{cat/sep}$ in full cells with LTO anode can be found in a similar way as for full cells with Gr anode by using Eq. 16 and the analytical expression of $L_{PZ}$ given in Supplementary Table S2, which is:

$$c_{cat/sep} \approx \begin{cases} \dfrac{6}{9-2f} c_0 & \text{UR-type cathode} \\ c_0 & \text{MZR-type cathode} \end{cases} \quad \text{S8)}$$

If replacing $c_{cat/sep}$ in Eq. 16 with $c_{cat/sep}(f=1)$ given by Eq. S8, we obtain an approximate expression of $L_{PZ}$ for full cells with LTO anode. It has the same form of Eq. 17 in the main text but with different $\gamma_f$ values: $\gamma_f = 12/7$ for UR-type cathodes and $\gamma_f = 1$ for MZR-type cathodes.

**Supplementary Note S3** — Selection criteria of experimental data presented in Figure 5c

The analytical model developed in this work assumes that the cathode is in fully deintercalated state at the beginning of the discharge process so that the reaction non-uniformity stems solely from electrolyte diffusion limitation. Among the experimental works we found in literature that systematically report the dependence of the discharge capacity on discharge rate and electrode thickness, many employed a symmetric cycling protocol, in which the discharge capacity was measured after battery cells were first charged at the same rate. At relatively high rates, this will cause the cathode particles to have a non-uniform initial SOC distribution at the beginning of discharge, which deviates from the model assumption. Furthermore, the discharge performance measured in symmetric cycling may be limited by the charging capacity preceding the discharge step, which causes the measurements to reflect the kinetic limitations during charging instead of discharging. As such, we include in Figure 5c only data from works that conducted asymmetric cycling tests, in which battery cells were first charged at a low C rate (<C/3) and then discharged at different rates. This ensures that the initial cell state is close to be fully charged prior discharging. We also limit the DoD range of the selected data to between 0.2 and 0.9 to ensure that they are in the electrolyte-limited regime.

**Supplementary Table S1.** Parameters used in P2D simulations (unless otherwise stated)

| Parameter | Symbol | Value | | | |
|---|---|---|---|---|---|
| Electrode properties | | | | | |
| | | NMC | LFP | Graphite | LTO |
| Cathode /anode particle radius (μm) | $r_{cat}/r_{an}$ | 1 | 0.1 | 1 | 0.1 |
| Cathode /anode porosity | $\epsilon_{cat}/\epsilon_{an}$ | 0.25 | 0.25 | 0.33 | 0.25 |
| Separator thickness (μm) | $L_{sep}$ | 25 | | | |
| Separator porosity | $\epsilon_{sep}$ | 0.55 | | | |
| Tortuosity | $\tau$ | $\tau = \epsilon^{-0.5}$ | | | |
| Maximum Li concentration in active materials (mol·m$^{-3}$) | $c_{S,max}$ | 49761 | 22806 | 31507 | 22800 |
| Initial concentration in active materials (mol·m$^{-3}$) | $c_{S0}$ | 22392 | 228 | 27411 | 19836 |
| Li diffusivity in active materials (m$^2$·s$^{-1}$) | $D_S$ | 10$^{-14}$ [ref. 1] | 10$^{-16}$ [ref. 2] | 9·10$^{-14}$ [ref. 3] | 10$^{-16}$ [ref. 4] |
| Electrode conductivity (S·m$^{-1}$) | $\sigma_S$ | 10 [ref. 5] | 10 [a] | 100 [ref. 3] | 100 [a] |
| Reaction rate constant (mol·m$^{-2}$·s$^{-1}$·(mol·m$^{-3}$)$^{-1.5}$) | $k_0$ | 3·10$^{-11}$ [ref. 5] | 3·10$^{-11}$ [a] | 3·10$^{-11}$ [ref. 2] | 10$^{-9}$ [ref. 6] |
| Charge transfer coefficient | $\alpha$ | 0.5 | | | |
| Exchange current density of Li anode (A·m$^{-2}$) | $i_0^{Li}$ | 20 [ref. 5] | | | |
| Equilibrium potential (V) | $U_{eq}$ | See note b | | | |
| Electrolyte (1M LiPF$_6$ in EC/DMC 50:50 wt.%) properties | | | | | |
| Initial salt concentration (mol·m$^{-3}$) | $c_0$ | 1000 [ref. 7] | | | |
| Transference number of cations | $t_+$ | 0.39 [ref. 7] | | | |
| Ambipolar diffusivity (m$^2$·s$^{-1}$) | $D_{amb}$ | 2.95·10$^{-10}$ [ref. 7] | | | |
| Concentration-dependent ionic conductivity (S·m$^{-1}$) | $\kappa(c)$ | $0.00233c$ [c] | | | |

Note:

**a.** Assumed.

**b.** The equilibrium potential profiles of NMC, LFP and LTO are extracted from Figure 2 in Ref. 8, Figure 2 in Ref. 9 and Figure 1 in Ref. 10. The equilibrium potential of graphite is adopted from Ref. 3 as

$$U_{eq,an}(x) = 0.124 + 1.5\exp(-70x) - 0.0351\tanh(\frac{x-0.286}{0.083}) - 0.0045\tanh(\frac{x-0.9}{0.119})$$

$$-0.035\tanh(\frac{x-0.99}{0.05}) - 0.0147\tanh(\frac{x-0.5}{0.034}) - 0.102\tanh(\frac{x-0.194}{0.142})$$

$$-0.022\tanh(\frac{x-0.98}{0.0164}) - 0.011\tanh(\frac{x-0.124}{0.0226}) + 0.0155\tanh(\frac{x-0.105}{0.029})$$

**c.** Calculated by $\kappa = F^2 D_{amb} c / \left(2RTt_+(1-t_+)\right)$

**Supplementary Table S2.** Analytical expressions of salt penetration depth $L_{PZ}$ in full cells with LTO anode

| | UR-type cathode | MZR-type cathode |
|---|---|---|
| $L_{PZ}$ | $\dfrac{-B + \sqrt{B^2 - 4AC}}{2A}$ | |
| $A$ | $\dfrac{1}{6}\tau_{cat} - \dfrac{(c_{S,max,cat} - c_{S0,cat})^2 (1-\epsilon_{cat})^2}{2\, c_{S0,an}{}^2 (1-\epsilon_{an})^2}\tau_{an}$ | $\dfrac{1}{2}\tau_{cat} - \dfrac{(c_{S,max,cat} - c_{S0,cat})^2 (1-\epsilon_{cat})^2}{2\, c_{S0,an}{}^2 (1-\epsilon_{an})^2}\tau_{an}$ |
| $B$ | $\dfrac{(c_{S,max,cat} - c_{S0,cat})(1-\epsilon_{cat})}{c_{S0,an}(1-\epsilon_{an})}\tau_{an}L_{an}$ $+ \dfrac{\tau_{cat}}{2\epsilon_{cat}}(\epsilon_{sep}L_{sep} + \epsilon_{an}L_{an})$ | $\dfrac{(c_{S,max,cat} - c_{S0,cat})(1-\epsilon_{cat})}{c_{S0,an}(1-\epsilon_{an})}\tau_{an}L_{an}$ $+ \dfrac{\tau_{cat}}{\epsilon_{cat}}(\epsilon_{sep}L_{sep} + \epsilon_{an}\, L_{an})$ |
| $C$ | $-\dfrac{FD_{amb}c_0}{I(1-t_+)}(\epsilon_{cat}L_{cat} + \epsilon_{sep}L_{sep} + \epsilon_{an}L_{an}) + \left(\dfrac{1}{2}L_{sep} + \dfrac{\epsilon_{an}}{\epsilon_{sep}}L_{an}\right)\tau_{sep}L_{sep}$ | |

**Supplementary Table S3.** Description and fitted scaling exponents of the experimental data in Figure 5c

| Label | Configuration | Active materials | Cathode thickness (μm) | C rate | Fitted $n$ | Reference |
|---|---|---|---|---|---|---|
| Yu-5%AB | Half cell | LFP | 45 – 120 | 0.5 – 8 | -0.832 | [11] |
| Yu-10%AB | Half cell | LFP | 40 – 125 | 1 – 8 | -0.952 | [11] |
| Zheng-NMC | Half cell | NMC (3:3:3) | 50 – 104 | 2 – 10 | -0.587 | [12] |
| Zheng-LFP | Half cell | LFP | 50 – 108 | 2 – 20 | -0.514 | [12] |
| Sander-homogenerous | Half cell | LCO | 310 | 0.2 – 1 | -0.619 | [13] |
| Sander-low tortuosity | Half cell | LCO | 310 | 0.5 – 2 | -0.714 | [13] |
| Bae | Half cell | LCO | 220 | 0.5 – 2 | -0.714 | [14] |
| Gallagher | Full cell | NMC (6:2:2) Graphite | 77 – 154 | 5 – 2 | -1.325 | [15] |
| Singh | Full cell | NMC(3:3:3) / Graphite | 255 – 305 | 0.33 – 0.5 | -1.807 | [16] |

**Supplementary Table S4.** Cell component properties used in calculating the cell-level specific capacity

| Parameter | Symbol | Value | | | |
|---|---|---|---|---|---|
| | | NMC | LFP | Lithium | Graphite |
| Density of cathode active material (g·cm$^{-3}$) | $\rho_{cat}$ | 4.77 | 3.6 | - | |
| Density of anode active material (g·cm$^{-3}$) | $\rho_{an}$ | - | - | 0.534 | 2.27 |
| Volume capacity of cathode active material (mAh·cm$^{-3}$) | $Q_V^0$ | 734 | 611 | - | - |
| Anode / cathode capacity ratio | $R_{an/cat}$ | - | - | 1.25 | 1.18 |
| Density of separator (g·cm$^{-3}$) | $\rho_{sep}$ | 0.946 | | | |
| Density of electrolyte (g·cm$^{-3}$) | $\rho_{electrolyte}$ | 1.3 | | | |
| Density of copper (g·cm$^{-3}$) | $\rho_{Cu}$ | 8.96 | | | |
| Density of aluminum (g·cm$^{-3}$) | $\rho_{Al}$ | 2.7 | | | |
| Current collector thickness (μm) | $L_{cc}$ | 15 (double-side coating assumed) | | | |

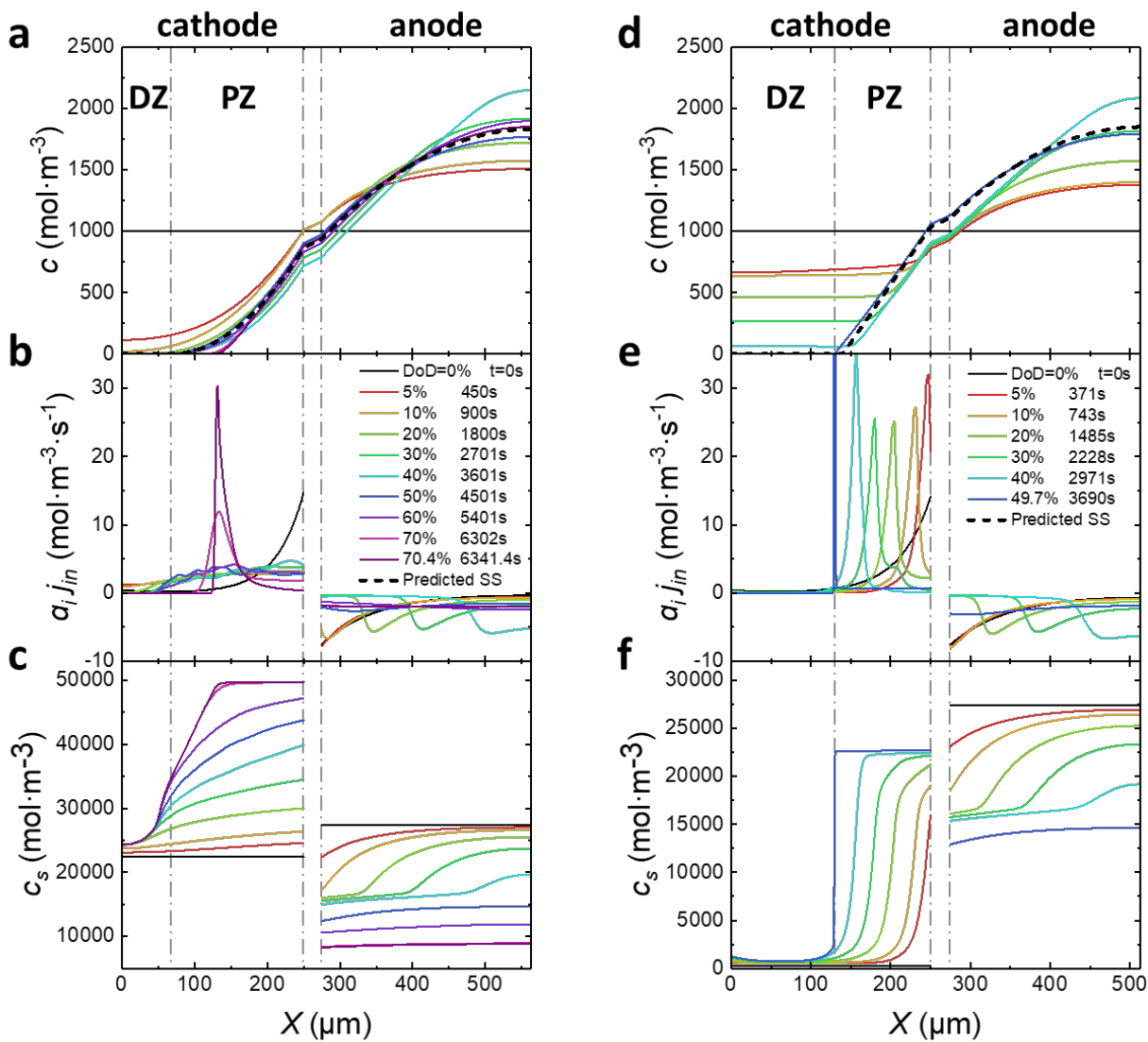

**Supplementary Figure S1. Discharge behavior of NMC/Gr and LFP/Gr full cells from P2D simulations.** Electrolyte is 1M LiPF$_6$ in EC/DMC (50:50 wt.%). $L_{cat}$ = 250 μm, $L_{an}$ is matched to $L_{cat}$ to give an anode/cathode capacity ratio of 1.18, and $I$ = 5.5 mA/cm$^2$. Other simulation parameters are given in Supplementary Table S1. The salt concentration $c$, reaction flux at the electrolyte/electrode interface $a_i j_{in}$ and the average Li concentration in cathode particles $c_s$ at different DoDs are shown in **a-c** for NMC/Gr full cell and **d-f** for LFP/Gr full cell, respectively. Cathode/current-collector interface is at $X$ = 0 μm. The dashed lines in **a** and **d** represent the (peudo-)steady-state salt distribution in electrolyte predicted by the analytical model.

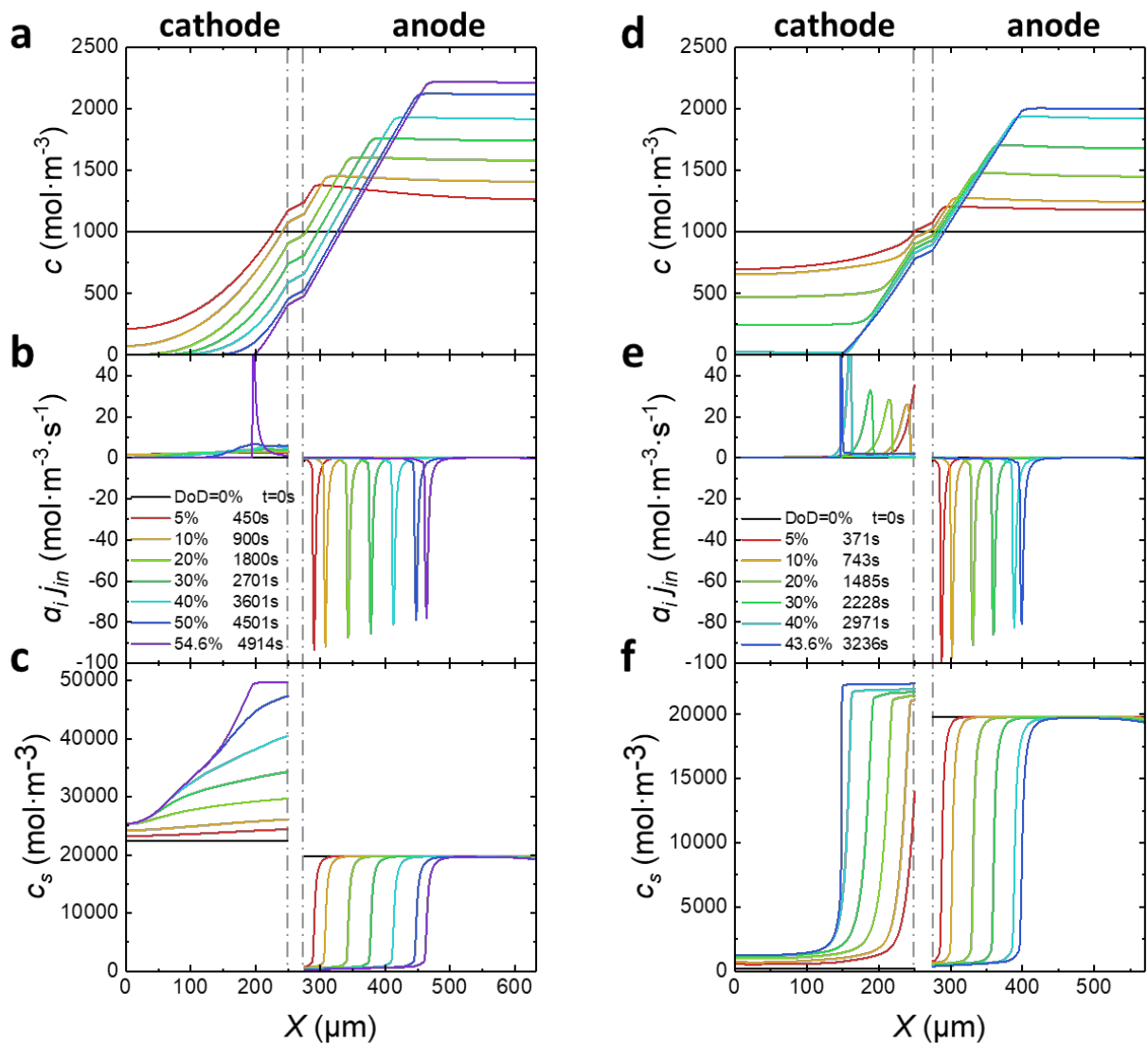

**Supplementary Figure S2. Discharge behavior of NMC/LTO and LFP/LTO full cells from P2D simulations.** Electrolyte is 1M $LiPF_6$ in EC/DMC (50:50 wt.%). $L_{cat}$ = 250 μm, $L_{an}$ is matched to $L_{cat}$ to give an anode/cathode capacity ratio of 1.18, and $I$ = 5.5 mA/cm². Other simulation parameters are given in Supplementary Table S1. The salt concentration $c$, reaction flux at the electrolyte/electrode interface $a_i j_{in}$ and the average Li concentration in cathode particles $c_s$ at different DoDs are shown in **a-c** for NMC/LTO full cell and **d-f** for LFP/LTO full cell, respectively. Cathode/current-collector interface is at $X$ = 0 μm.

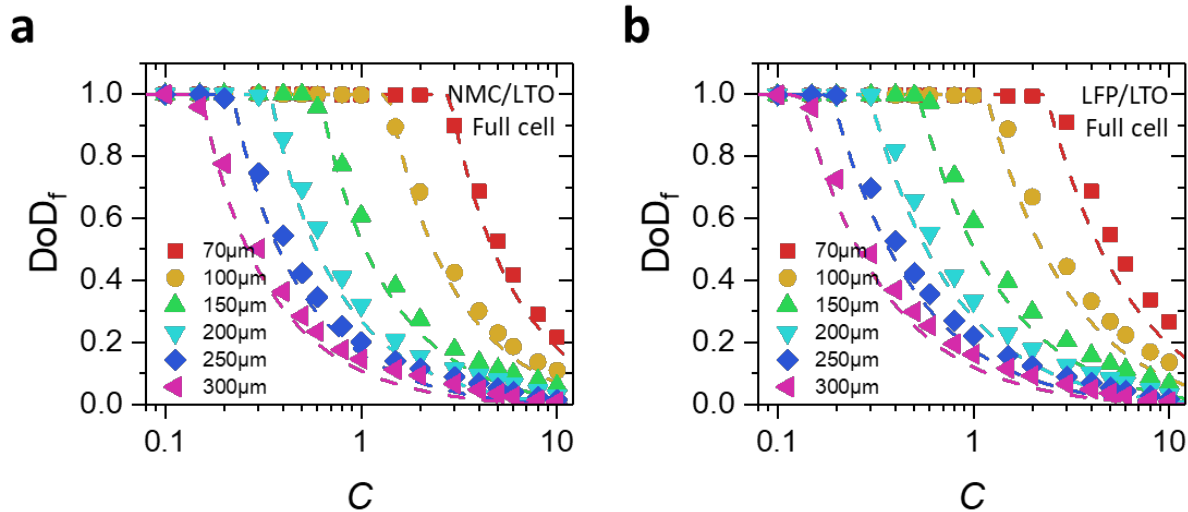

**Supplementary Figure S3. Rate performance of NMC/LTO and LFP/LTO full cells predicted by P2D simulations (solid symbols) vs. the analytical model (dashed lines). a** NMC/LTO full cells. **b** LFP/LTO full cells. Different symbols and line colors represent different $L_{cat}$, which varies between 70 μm and 300 μm. $L_{an}$ is matched to $L_{cat}$ to give an anode/cathode capacity ratio of 1.18. Other simulation parameters are listed in Supplementary Table S1.

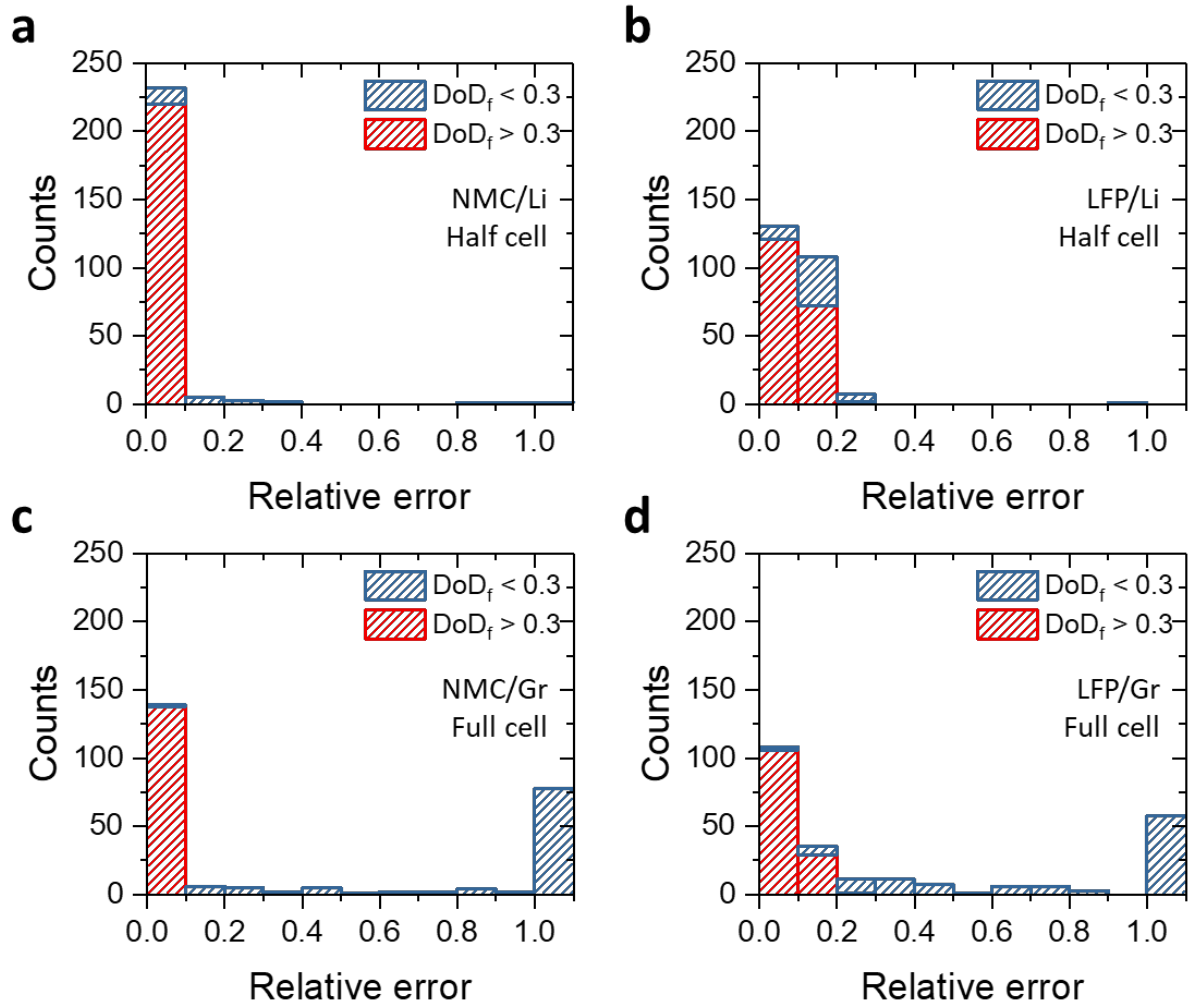

**Supplementary Figure S4. Relative error of the discharge capacity predicted by the analytical model compared to P2D simulations.** Analysis is done on the 246 simulations included in Figure 4 **a.** NMC half cell. **b.** LFP half cell. **c**. NMC/Gr full cell. **d**. LFP/Gr full cell. Red and blue bars represent counts of simulations with $DoD_f > 0.3$ and $DoD_f < 0.3$, respectively. The counts at the relative error = 1 in full cells (**c**, **d**) correspond to the cases for which the analytical model predicts negative $L_{PZ}$ and $DoD_f = 0$ is set by Eq. 15.

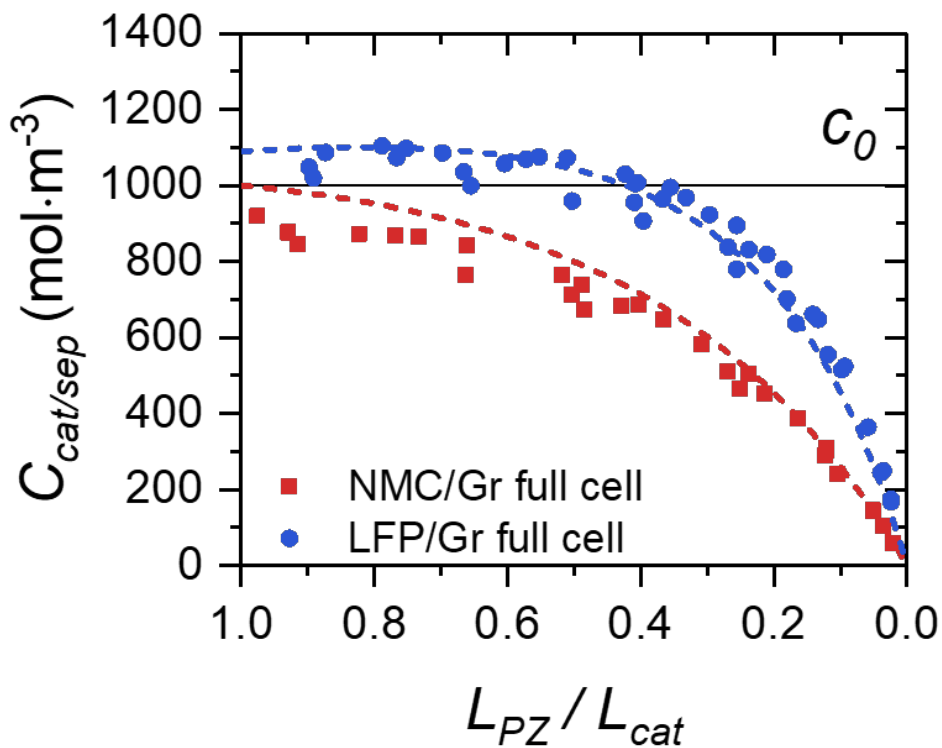

**Supplementary Figure S5. Relation between the (pseudo-)steady-state salt concentration at the cathode/separator interface, $c_{cat/sep}$, and $L_{PZ}/L_{cat}$ for NMC/Gr (■) and LFP/Gr (●) full cells.** Data are taken from the cases presented in Figure 3c,d. The dashed lines are estimates given by Eq. S7.

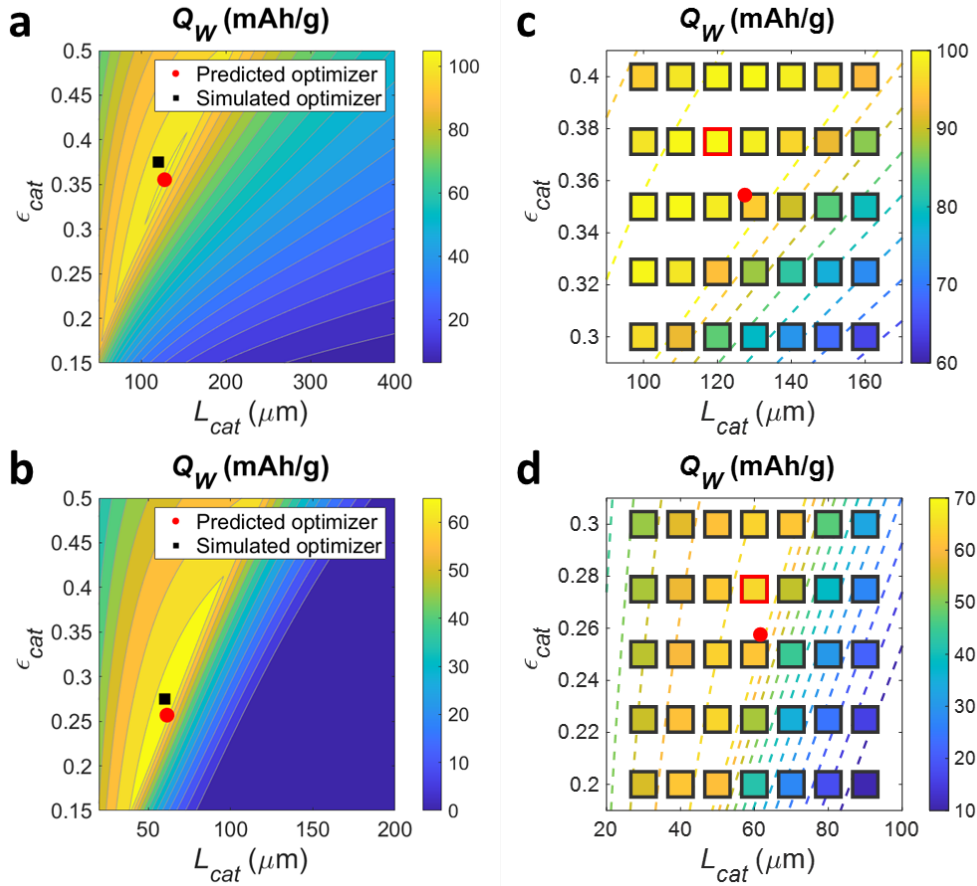

**Supplementary Figure S6. Optimization of NMC half cells and NMC/Gr full cells at 5C discharge.** **a and b**, contour plots of cell-level specific capacity $Q_w$ at 5C discharge as a function of $L_{cat}$ and $\epsilon_{cat}$ for NMC half and NMC/Gr full cells, respectively. The anode/cathode capacity ratio is held constant at 1.18, and $L_{an}$ and $\epsilon_{an}$ vary in proportion to $L_{cat}$ and $\epsilon_{cat}$, respectively. The red circle and black square represent the global optimum ($L_{cat}^{opt}, \epsilon_{cat}^{opt}$) predicted by the model and P2D simulations, respectively. **c and d**, close-up of the neighborhood of ($L_{cat}^{opt}, \epsilon_{cat}^{opt}$) for 5C discharge in the parameter space for NMC half and NMC/Gr full cells, respectively. The filled color of square symbols represents simulated $Q_w$ at different $L_{cat}$ and $\epsilon_{cat}$. The square symbol with red edges represents ($L_{cat}^{opt}, \epsilon_{cat}^{opt}$) located by P2D simulations, and the red circle is the model prediction.